\PassOptionsToPackage{unicode}{hyperref}
\PassOptionsToPackage{hyphens}{url}
\documentclass[
]{article}
\usepackage{amsmath,amssymb}
\usepackage{iftex}
\ifPDFTeX
  \usepackage[T1]{fontenc}
  \usepackage[utf8]{inputenc}
  \usepackage{textcomp} 
\else 
  \usepackage{unicode-math} 
  \defaultfontfeatures{Scale=MatchLowercase}
  \defaultfontfeatures[\rmfamily]{Ligatures=TeX,Scale=1}
\fi
\usepackage{lmodern}
\ifPDFTeX\else
\fi
\IfFileExists{upquote.sty}{\usepackage{upquote}}{}
\IfFileExists{microtype.sty}{
  \usepackage[]{microtype}
  \UseMicrotypeSet[protrusion]{basicmath} 
}{}
\makeatletter
\@ifundefined{KOMAClassName}{
  \IfFileExists{parskip.sty}{%
    \usepackage{parskip}
  }{
    \setlength{\parindent}{0pt}
    \setlength{\parskip}{6pt plus 2pt minus 1pt}}
}{
  \KOMAoptions{parskip=half}}
\makeatother
\usepackage{xcolor}
\usepackage{longtable,booktabs,array}
\usepackage{calc} 
\usepackage{etoolbox}
\makeatletter
\patchcmd\longtable{\par}{\if@noskipsec\mbox{}\fi\par}{}{}
\makeatother
\IfFileExists{footnotehyper.sty}{\usepackage{footnotehyper}}{\usepackage{footnote}}
\makesavenoteenv{longtable}
\usepackage{graphicx}
\makeatletter
\def\maxwidth{\ifdim\Gin@nat@width>\linewidth\linewidth\else\Gin@nat@width\fi}
\def\maxheight{\ifdim\Gin@nat@height>\textheight\textheight\else\Gin@nat@height\fi}
\makeatother
\setkeys{Gin}{width=\maxwidth,height=\maxheight,keepaspectratio}
\makeatletter
\def\fps@figure{htbp}
\makeatother
\providecommand{\tightlist}{%
  \setlength{\itemsep}{0pt}\setlength{\parskip}{0pt}}
\usepackage{etoolbox}
\AtBeginEnvironment{longtable}{\footnotesize}
\usepackage{float}
\makeatletter
\def\fps@figure{H}
\makeatother
\usepackage{newunicodechar}
\usepackage{amssymb} 

\newunicodechar{χ}{\ensuremath{\chi}}
\newunicodechar{θ}{\ensuremath{\theta}}
\newunicodechar{λ}{\ensuremath{\lambda}}
\newunicodechar{α}{\ensuremath{\alpha}}
\newunicodechar{κ}{\ensuremath{\kappa}}
\newunicodechar{Δ}{\ensuremath{\Delta}}

\newunicodechar{→}{\ensuremath{\rightarrow}}
\newunicodechar{↓}{\ensuremath{\downarrow}}
\newunicodechar{↑}{\ensuremath{\uparrow}}
\newunicodechar{↔}{\ensuremath{\leftrightarrow}}

\newunicodechar{≈}{\ensuremath{\approx}}
\newunicodechar{≥}{\ensuremath{\geq}}
\newunicodechar{≤}{\ensuremath{\leq}}
\newunicodechar{≠}{\ensuremath{\neq}}
\newunicodechar{∼}{\ensuremath{\sim}}
\newunicodechar{∞}{\ensuremath{\infty}}
\newunicodechar{±}{\ensuremath{\pm}}
\newunicodechar{×}{\ensuremath{\times}}
\newunicodechar{−}{\ensuremath{-}}

\newunicodechar{✓}{\ensuremath{\checkmark}}
\newunicodechar{✅}{\ensuremath{\checkmark}}
\newunicodechar{❌}{\ensuremath{\times}}

\newunicodechar{§}{\S}

\newunicodechar{⁰}{\textsuperscript{0}}
\newunicodechar{¹}{\textsuperscript{1}}
\newunicodechar{²}{\textsuperscript{2}}
\newunicodechar{³}{\textsuperscript{3}}
\newunicodechar{⁴}{\textsuperscript{4}}
\newunicodechar{⁵}{\textsuperscript{5}}
\newunicodechar{⁶}{\textsuperscript{6}}
\newunicodechar{⁷}{\textsuperscript{7}}
\newunicodechar{⁸}{\textsuperscript{8}}
\newunicodechar{⁹}{\textsuperscript{9}}
\newunicodechar{⁻}{\textsuperscript{\ensuremath{-}}}

\newunicodechar{₀}{\textsubscript{0}}
\ifLuaTeX
  \usepackage{selnolig}  
\fi
\IfFileExists{bookmark.sty}{\usepackage{bookmark}}{\usepackage{hyperref}}
\IfFileExists{xurl.sty}{\usepackage{xurl}}{} 
\hypersetup{
  hidelinks,
  pdfcreator={LaTeX via pandoc}}

\author{}
\date{}

\begin{document}

\hypertarget{code-lifespan-survival-analysis-clsa-predicting-the-survival-of-source-code-lines-using-ast-aware-mining}{%
\section{Code Lifespan Survival Analysis (CLSA): Predicting the Survival
of Source Code Lines Using AST-Aware
Mining}\label{code-lifespan-survival-analysis-clsa-predicting-the-survival-of-source-code-lines-using-ast-aware-mining}}

\textbf{Authors}: Pavel Gurov \textbf{Affiliations}: Independent
Researcher

\noindent\fbox{\parbox{\dimexpr\linewidth-2\fboxsep-2\fboxrule\relax}{%
\textbf{This preprint is superseded.} The analysis was redone after version 4, and several results below
should no longer be cited. The median survival of about 3.7 years describes a world without rewrites;
following lines through rewrites gives 3.0 years. The ``stabilize-or-die'' phase at 95.7 days is a median
among dead lines: the risk of deletion falls steadily with age, with no threshold. The effects of night-time
commits and of conditional context on removal are not significant once standard errors are clustered by
repository. See P.~Gurov, ``Mismeasuring the Life of a Line of Code: Lessons Learned the Hard Way'',
\href{https://arxiv.org/abs/2611.10530}{arXiv:2611.10530}. The text below is unchanged from version 4.}}

\hypertarget{abstract}{%
\subsection{Abstract}\label{abstract}}

\textbf{Context:} Predicting which source code lines will be deleted ---
and when --- has direct implications for maintenance planning,
technical-debt management, and code-review prioritization. Existing MSR
approaches operate at file or method granularity, masking the
heterogeneous deletion risk of individual statements.

\textbf{Objective:} We introduce \textbf{Code Lifespan Survival Analysis
(CLSA)}, the first framework to model the deletion risk of individual
source lines \emph{from covariates} --- where prior line-granularity
work estimated lifespans but found no significant predictors. CLSA
treats each source line as a right-censored survival subject and
estimates its deletion risk from structural, contextual, and temporal
covariates, with its strongest predictors computable statically from a
single source file --- combining AST structure with lexical features
such as line token count --- without any version history or historical
bug data.

\textbf{Method:} We mine 32.5 million line birth events from 120 active
open-source TypeScript repositories. A 5-stage bipartite matching
pipeline separates true deletions from refactoring noise (migrations and
semantic rewrites), preventing 8.3 million false death classifications
--- a prerequisite for unbiased line-level survival estimation. We fit a
Cox Proportional Hazards model with 15 covariates and assess robustness
via Weibull and Log-Logistic AFT models, a shared gamma frailty model,
time-stratified landmark analysis, and four sensitivity subsets.

\textbf{Results:} The Kaplan--Meier median survival time is 1,335--1,352
days (about 3.7 years, consistent across the analytical sample, the full
32.5-million-line population, and an independent resample); among
deleted lines, the median lifespan is 95.7 days, marking an early
``stabilize-or-die'' phase. Covariate effects are time-varying for some,
but not all, covariates. The clearest time dynamic is contextual: lines
in conditional branches show no detectable effect at birth (HR = 0.98,
0--90 days, not significant) but become a risk factor after 90 days (HR
= 1.23) --- mechanistically explaining their proportional-hazards
violation. Line token count is a steady protective lexical factor (HR =
0.87 in the main model) whose effect is essentially flat with age (HR
0.89--0.91 across the three regimes, with no proportional-hazards
violation). Repository identity is the single largest predictive factor:
a shared gamma frailty model (variance θ = 1.451) raises concordance
from 0.587 to 0.676, outweighing every structural covariate.

\textbf{Conclusion:} Line-level survival modeling is tractable and
yields interpretable, largely statically-computable risk signals (from
AST structure and lexical features) for software evolution. The
three-regime time structure and the dominance of repository frailty
provide both an empirical basis and a calibration recipe for
time-conditional risk scoring in IDEs and code review.

\begin{center}\rule{0.5\linewidth}{0.5pt}\end{center}

\hypertarget{introduction}{%
\subsection{1. Introduction}\label{introduction}}

Source code is inherently dynamic. Across a project's lifecycle, lines
of code are continuously added, modified, migrated between files, and
deleted---processes that collectively constitute software evolution.
While traditional metrics in Mining Software Repositories (MSR)---such
as code churn {[}8{]}, file age, or architectural ownership---provide
valuable macro-level insights, they often lack the granularity required
to evaluate individual statement volatility. File-level and method-level
aggregations mask the heterogeneous risk profiles of individual
statements: a stable function signature and a volatile expression within
the same method face fundamentally different hazard dynamics.

In this work, we model each source code line as a survival subject---an
entity born at the commit where it first appears and at risk of
permanent deletion throughout the repository's history. Leveraging
classical survival analysis techniques from actuarial science and
biostatistics, we estimate the probability that a given line persists in
the codebase over time, conditional on its structural and contextual
covariates. Crucially, our framework distinguishes three change
topologies: \emph{migrations} (lines relocated across files),
\emph{modifications} (semantically rewritten lines matched via composite
Sørensen--Dice + Ratcliff/Obershelp similarity), and \emph{true deaths}
(permanent deletions without a semantic successor). Only true deaths
constitute the event of interest; migrations and modifications are
treated as censored observations, preserving the line's survival
identity.

Empirically, we analyze 32.5 million line birth events from 120 active
TypeScript repositories on GitHub, building a deterministic sample of
350,000 lines for multivariate modeling. TypeScript was selected for its
rich AST node taxonomy and the large number of active open-source
projects it hosts, providing a representative spectrum of domains and
project scales.

Our main contributions are as follows:

\begin{enumerate}
\def\labelenumi{\arabic{enumi}.}
\item
  \textbf{5-Stage Semantic Alignment Pipeline}: We introduce a bipartite
  matching algorithm that isolates true line deletions from refactoring
  noise through five progressive stages: intra-file exact structural
  matching, global AST-equivalent detection, Hungarian algorithm--based
  lexico-semantic assignment {[}2{]}, cross-file similarity matching,
  and high-confidence AST-agnostic intra-file matching. The pipeline
  prevents 8.3 million false deaths in the full population --- a
  prerequisite for unbiased survival estimation at line granularity.
\item
  \textbf{Time-Stratified Landmark Analysis}: We fit separate Cox models
  on three landmark time bands (\textless{} 90 days, 90--365 days, 365+
  days) to resolve which covariate effects are genuinely time-varying.
  The sharpest dynamic is contextual: \texttt{in\_condition} shows no
  detectable effect in new code (HR = 0.98, not significant) but becomes
  a risk factor after 90 days (HR = 1.23). Line token count, by
  contrast, is essentially flat across all three regimes (HR 0.89--0.91)
  and does not violate the proportional-hazards assumption. These
  findings show that PH violations flag which covariates are genuinely
  time-varying, rather than being a uniform statistical nuisance.
\item
  \textbf{Repository Frailty Quantification}: We fit a shared gamma
  frailty Cox model (R \texttt{survival} package) and estimate frailty
  variance θ = 1.451 across the analyzed repositories (clustering
  restricted to those with ≥ 30 events; LRT χ² = 41,571, p ≈ 0),
  confirming that repository-level heterogeneity is the dominant source
  of survival variance and improving C-index from 0.587 to 0.676.
\item
  \textbf{Covariate-Effect Modeling at Line Granularity}: Prior
  line-level survival work measured \emph{how long} lines live but
  reported no covariates that significantly influence longevity
  (Spinellis et al.~{[}25{]}, via per-project OLS on lifespan). We are
  the first to recover \emph{why} lines live or die at this granularity
  --- fitting proportional-hazards, frailty, and AFT models with
  explicit right-censoring to obtain per-statement hazard estimates
  computable statically from a single file snapshot (from AST structure
  and lexical features such as line token count), without historical bug
  data. We show that \texttt{ast\_group\_expression} and
  \texttt{in\_function} are the most stable structural predictors across
  all sensitivity models.
\item
  \textbf{AST-Aware Structural Context Features}: We introduce two
  binary AST-derived covariates --- \texttt{in\_condition} and
  \texttt{in\_function} --- whose survival effects are robust across
  marginal Cox, frailty Cox, AFT, and time-stratified models.
  \texttt{in\_loop} is shown to be a repository-level proxy rather than
  a genuine line-level predictor, a distinction requiring the frailty
  model to detect {[}1{]}; \texttt{in\_try\_catch} attenuates
  substantially under the same conditioning but remains statistically
  significant.
\end{enumerate}

The remainder of the paper is organized as follows. §2 reviews related
work. §3 formalizes the survival problem. §4 states research questions.
§5 describes the data collection and statistical methodology, including
the three-regime time-stratified design. §6 presents empirical results.
§7 discusses practical implications. §8 addresses threats to validity.
§9 concludes.

\begin{center}\rule{0.5\linewidth}{0.5pt}\end{center}

\hypertarget{background-and-related-work}{%
\subsection{2. Background and Related
Work}\label{background-and-related-work}}

\hypertarget{mining-software-repositories-code-change-analysis}{%
\subsubsection{2.1 Mining Software Repositories \& Code Change
Analysis}\label{mining-software-repositories-code-change-analysis}}

The MSR community has extensively studied code churn and file-level
evolution to inform defect prediction {[}6{]} and refactoring scheduling
{[}7{]}. Nagappan and Ball {[}8{]} demonstrated that code churn metrics
are strong predictors of defect density. Hassan {[}9{]} introduced the
concept of ``entropy of changes'' as a complexity proxy. However, most
existing works aggregate changes at the file or method level using
traditional \texttt{git\ diff} heuristics. These macro-approaches often
mask the underlying statement-level risk dynamics. Tools like PyDriller
{[}4{]} have revolutionized commit traversal, but moving down to
accurate line-level tracking requires mitigating the severe noise
inherent to spatial line shifting. Hattori and Lanza {[}10{]}
categorized commit types but did not model individual line survival.

The most influential line-level blame technique in MSR is the
\textbf{SZZ algorithm} {[}22{]} (Śliwerski, Zimmermann, and Zeller,
2005), which propagates a bug-fix commit backward through
\texttt{git\ blame} to identify the change that \emph{introduced} the
fault. Kim et al.~{[}23{]} later automated and scaled SZZ across large
repositories. CLSA differs from SZZ in three fundamental respects: (i)
\emph{direction} --- SZZ traces history backward from a known defect;
CLSA models the forward survival trajectory of every line from birth
onward; (ii) \emph{trigger} --- SZZ is event-driven (applied only when a
bug fix is committed); CLSA is population-wide (applied continuously,
for every line); (iii) \emph{objective} --- SZZ produces a deterministic
fault-origin attribution; CLSA produces a probabilistic survival
estimate. Where SZZ asks ``who introduced this bug?'', CLSA asks ``how
long will this line exist?'' Both approaches depend critically on
accurate line-level identity tracking across commits --- the challenge
addressed by our 5-stage matching pipeline (§5.2).

\hypertarget{survival-analysis-in-software-engineering}{%
\subsubsection{2.2 Survival Analysis in Software
Engineering}\label{survival-analysis-in-software-engineering}}

Survival analysis encompasses statistical procedures where the outcome
variable of interest is the time until an event occurs. Samoladas et
al.~{[}3{]} applied survival analysis to study the longevity of
open-source projects, demonstrating the utility of Kaplan--Meier and Cox
models in SE contexts. Shahzad et al.~{[}11{]} used survival models to
analyze vulnerability life cycles in software systems. Raemaekers et
al.~{[}5{]} applied similar techniques to study semantic versioning
compliance and breaking changes in Maven libraries. Bavota et
al.~{[}12{]} empirically studied how test smells affect maintenance
effort and test-suite quality across software evolution, documenting
longitudinal change patterns in open-source test code. Beyond projects,
time-to-event methods have been applied across the SE granularity
spectrum: to software-project duration {[}29{]}, to source-code survival
via the Kaplan--Meier estimator {[}30{]}, to the longevity of Debian
packages with unresolved conflicts {[}31{]}, and to the usage lifetime
of database frameworks in Java projects {[}32{]}. This work descends
from the older theory of software \emph{aging} and \emph{decay} ---
Lehman's laws of software evolution {[}26{]}, Parnas's notion of
software aging {[}27{]}, and Eick et al.'s empirical demonstration that
code decays as change accumulates {[}28{]} --- and connects to
complementary evidence that older code is more stable: complex-network
analyses of software find age-dependent attachment in which older
elements are less volatile {[}33{]}, while structural and configuration
code smells have been associated with reduced evolvability and higher
maintenance risk {[}34{]}, {[}35{]}. Crucially, survival analysis
naturally handles \emph{right-censored} data---cases where the study
ends before the outcome event occurs (e.g., active code still present at
the time of the latest commit). Correct handling of censored
observations is essential: treating all lines as ``dead'' would severely
bias survival estimates downward and produce misleading hazard ratios
{[}13{]}. Most of these studies operate at a coarse unit --- the
open-source project {[}3{]}, the library or API {[}5{]}, the
vulnerability {[}11{]}, or the class {[}17{]}. The one prior study that
operates at our granularity is Spinellis et al.~{[}25{]}, who tracked
the birth and death of 3.3 billion individual source-code lines and
tokens across 89 long-lived repositories and estimated their lifespans
with Kaplan--Meier and Weibull models --- the same non-parametric and
parametric tooling we adopt. Their results establish the baseline
phenomenon CLSA builds on: code lines are durable (median lifespan ≈ 2.4
years) and exhibit \emph{infant mortality}, with a Weibull shape
parameter below one in every project, so that young lines face the
highest deletion hazard. Crucially, however, when they searched for
\emph{what} makes a line long- or short-lived, they reported a null
result: a per-project linear regression of lifespan on line-level
features (length, indentation, brackets, commas, comments, and the like)
yielded R² \textless{} 0.1 for almost every project, token type was
non-discriminative, and developer experience, project size, and
programming language showed no consistent effect. They concluded that
they could not determine factors that significantly influence line or
token longevity.

This null result motivates our central methodological choice. CLSA
differs from {[}25{]} not in the unit of analysis --- both model the
individual line --- but in the inferential machinery applied to it.
Regressing a censored lifespan directly with ordinary least squares
discards the survivors (more than half of all lines in our data are
never deleted) and cannot separate a covariate's effect on the
\emph{timing} of deletion from its effect on the \emph{event} itself. By
instead fitting Cox proportional-hazards, gamma-frailty, and
accelerated-failure-time models with explicit right-censoring and
Benjamini--Hochberg FDR control, we recover stable, interpretable
covariate effects --- most prominently a protective effect of line
content (token count) and of nesting depth --- that an OLS-on-lifespan
analysis leaves invisible. Where prior line-level work measured
\emph{how long} lines live, CLSA models \emph{why}, turning a reported
absence of effects into a structured, statically computable hazard
profile.

\hypertarget{ast-based-code-analysis}{%
\subsubsection{2.3 AST-Based Code
Analysis}\label{ast-based-code-analysis}}

Recent advances in tree-sitter and similar incremental parsers allow for
structural, rather than purely textual, analysis of code. GumTree
{[}1{]} introduced fine-grained AST differencing that maps individual
node-level changes, enabling precise classification of code
modifications. Fluri et al.~{[}14{]} proposed ChangeDistiller, which
extracts fine-grained source code changes from AST diffs. Combining
AST-derived structural features with revision history is known to be
superior to simple grep-based heuristics {[}1{]}, allowing us to
categorize nodes directly from the syntactic topology. Our approach
extends this line of work by using tree-sitter AST node types as
survival covariates.

\hypertarget{code-complexity-and-change-proneness}{%
\subsubsection{2.4 Code Complexity and
Change-Proneness}\label{code-complexity-and-change-proneness}}

Several studies have linked code complexity metrics to change frequency.
Gil and Lalouche {[}15{]} comprehensively evaluated complexity metrics,
finding that cyclomatic complexity and nesting depth correlate with
change-proneness. Munson and Elbaum {[}16{]} showed that code complexity
metrics serve as predictors of fault-prone modules. Khomh et
al.~{[}17{]} applied survival analysis specifically to study the
relationship between code smells and change-proneness, finding that
classes with certain anti-patterns exhibit higher hazard rates---a
finding conceptually parallel to our line-level analysis.

\hypertarget{technical-debt-and-code-decay}{%
\subsubsection{2.5 Technical Debt and Code
Decay}\label{technical-debt-and-code-decay}}

The concept of ``code decay'' has been studied through the lens of
technical debt {[}18{]}, and behavioral code analysis using version
control history has been proposed to identify high-risk code regions
{[}19{]}. Our work complements this perspective by directly modeling the
time-to-event dynamics of individual code lines, providing quantitative
survival estimates rather than qualitative risk labels.

\begin{center}\rule{0.5\linewidth}{0.5pt}\end{center}

\hypertarget{problem-definition}{%
\subsection{3. Problem Definition}\label{problem-definition}}

\hypertarget{line-as-a-survival-subject}{%
\subsubsection{3.1 Line as a Survival
Subject}\label{line-as-a-survival-subject}}

We rigorously define each source code line \(i\) as a random variable
experiencing time-to-event dynamics:

\begin{itemize}
\tightlist
\item
  \textbf{Birth time (\(t_i^b\))}: The commit timestamp where line \(i\)
  first naturally appears.
\item
  \textbf{Censoring time (\(C_i\))}: The timestamp of the last commit
  observed in the repository containing line \(i\).
\item
  \textbf{Death time (\(t_i^d\))}: The timestamp of the commit in which
  line \(i\) is permanently removed from the codebase. This quantity is
  defined only for lines whose removal is actually observed within the
  observation window, i.e.~when \(t_i^b < t_i^d \le C_i\).
\item
  \textbf{Event indicator}: \(\delta_i = 1\) if a true-death event
  (§3.2) is observed for line \(i\) at or before \(C_i\), and
  \(\delta_i = 0\) otherwise.
\item
  \textbf{Observed duration}:
\end{itemize}

\[T_i = \begin{cases} t_i^d - t_i^b, & \delta_i = 1 \\ C_i - t_i^b, & \delta_i = 0 \end{cases}\]

Because \(t_i^d\) is left undefined for lines that are never removed,
the duration is stated case-wise rather than as a minimum over the two
quantities; \(T_i > 0\) holds by construction in both branches.

Lines without an explicitly observed death are right-censored. In our
dataset, 66.1\% of observations are right-censored, indicating that the
majority of code lines remain alive at the end of the observation
period.

\hypertarget{event-types-and-change-topologies}{%
\subsubsection{3.2 Event Types and Change
Topologies}\label{event-types-and-change-topologies}}

A naïve \texttt{git\ diff}-based approach treats every textual deletion
as a death event, severely inflating hazard estimates. To mitigate this
refactoring noise, we classify every observed change into one of three
topologies:

\begin{itemize}
\tightlist
\item
  \textbf{Migration}: A line disappears from its original file path but
  reappears with identical content and matching AST node type in another
  file (or at a different location after a file rename). The line
  retains its original identity (UUID) and continues to accumulate
  survival time. \emph{Migration ≠ death.}
\item
  \textbf{Modification}: A line is textually altered but remains
  semantically continuous---matched via composite similarity
  (Sørensen--Dice + Ratcliff/Obershelp, threshold ≥ 0.6) with the same
  AST node type. A new identity (UUID) is created and linked to the
  original via an evolution record (\texttt{line\_evolution} table). The
  original line is treated as \emph{right-censored} at the modification
  commit, not as dead. The new identity begins its own independent
  survival trajectory.
\item
  \textbf{True Death}: A line is permanently removed from the codebase
  without any semantic successor in the same commit. Only hard deletions
  (\texttt{hard\_delete}) and file-level deletions
  (\texttt{file\_delete}) constitute the event of interest
  (\(\delta_i = 1\)).
\end{itemize}

This three-way classification ensures that the survival model captures
genuine code extinction rather than refactoring artifacts {[}2{]}.

\begin{center}\rule{0.5\linewidth}{0.5pt}\end{center}

\hypertarget{research-questions}{%
\subsection{4. Research Questions}\label{research-questions}}

To structure our investigation, we formulate the following research
questions:

\begin{itemize}
\tightlist
\item
  \textbf{RQ1 (Baseline Survival):} What is the baseline survival
  distribution of source code lines in active TypeScript projects?
\item
  \textbf{RQ2 (Syntax Role):} How does the AST role of a line (e.g.,
  declaration vs.~expression) affect its hazard rate?
\item
  \textbf{RQ3 (Structural Complexity):} How does structural complexity
  (such as nesting depth and enclosing scope type) influence hazard
  rates?
\item
  \textbf{RQ4 (Temporal Context):} Does the time of writing (e.g.,
  weekend vs.~weekday, day vs.~night) affect the stability of the code?
\item
  \textbf{RQ5 (Global Uniqueness):} How does the global uniqueness
  (LIDF) of a line correlate with its stability?
\item
  \textbf{RQ6 (Change-Type Awareness):} How do different change types
  (hard deletion vs.~semantic rewrite vs.~migration) influence observed
  survival behavior?
\end{itemize}

\begin{center}\rule{0.5\linewidth}{0.5pt}\end{center}

\hypertarget{data-collection-and-analytical-methodology}{%
\subsection{5. Data Collection and Analytical
Methodology}\label{data-collection-and-analytical-methodology}}

The core challenge of MSR in time-to-event analysis is false positives:
a line merely changing its indentation or moving to another file appears
as a standard Git deletion, skewing survival metrics heavily downward.
To solve this, we implemented a sophisticated analytical framework.

\hypertarget{repository-extraction-and-syntactic-filtering}{%
\subsubsection{5.1 Repository Extraction and Syntactic
Filtering}\label{repository-extraction-and-syntactic-filtering}}

Using PyDriller {[}4{]}, we iterate over the repository's entire commit
history, restricting to \texttt{.ts} and \texttt{.tsx} file types.
\textbf{Generated and vendor files are excluded before any line-level
processing}: paths containing the segments \texttt{dist/},
\texttt{build/}, \texttt{.next/}, \texttt{out/}, or
\texttt{node\_modules/}, TypeScript declaration files (\texttt{.d.ts}),
and files whose first non-empty line begins with a standard
generated-code marker (\texttt{//\ @generated},
\texttt{//\ Code\ generated}, \texttt{/*\ eslint-disable\ */}) are
skipped entirely. For each remaining modified file, we parse both
\texttt{source\_code\_before} and \texttt{source\_code} with
tree-sitter, enabling precise mapping of diff line numbers to their AST
nodes. Two categories of lines are additionally excluded from the event
pool:

\begin{enumerate}
\def\labelenumi{\arabic{enumi}.}
\tightlist
\item
  \textbf{Comment lines}: Any line whose AST node is classified as a
  \texttt{comment} by tree-sitter is discarded. This is an
  AST-structural filter, not a regex heuristic, ensuring robust
  detection of single-line (\texttt{//}), block (\texttt{/*\ */}), and
  JSDoc comments.
\item
  \textbf{Trivial tokens} (length \textless{} 5 characters): Lines
  consisting solely of structural punctuation---braces (\texttt{\{},
  \texttt{\}}), parentheses, semicolons, or short keywords---are
  excluded. The 5-character threshold was chosen empirically to filter
  syntactic scaffolding (e.g., \texttt{\});}, \texttt{else},
  \texttt{\}\ else\ \{}) while retaining the shortest meaningful
  statements (e.g., \texttt{i++;}, \texttt{break;}, \texttt{x\ =\ 0}).
  These trivial tokens carry no semantic information relevant to
  survival analysis and would otherwise dominate the dataset with
  near-infinite lifespans, biasing censoring rates upward.
\end{enumerate}

\hypertarget{multi-stage-optimal-alignment-pipeline}{%
\subsubsection{5.2 Multi-Stage Optimal Alignment
Pipeline}\label{multi-stage-optimal-alignment-pipeline}}

Lines surviving the initial AST filter undergo string cleaning
(whitespace stripping). Additions and deletions are then pooled into
bipartite sets. To resolve migrations and semantic rewrites from true
deaths, we execute a rigorous 5-stage matching pipeline:

\begin{enumerate}
\def\labelenumi{\arabic{enumi}.}
\tightlist
\item
  \textbf{Exact Structural Match (Intra-File):} Exact string content
  with matching AST node type within the same file (or its rename
  target). Handles the common case of pure line relocation or unchanged
  code in a modified file.
\item
  \textbf{Global Structural Match:} Exact string and AST-type
  equivalents migrating to any other file across the repository.
  Captures full-file moves and directory-level restructuring not caught
  by Git rename detection.
\item
  \textbf{Linear Sum Assignment (Intra-File, AST-aware):} For remaining
  unmatched lines within the same file path, we construct a cost matrix
  comparing all remaining unassigned additions to deletions. We apply
  the Hungarian minimum-cost assignment algorithm {[}24{]} over a
  composite similarity matrix. The similarity score is a weighted
  combination of the Sørensen--Dice coefficient on tokenized strings
  (70\%) and the Ratcliff/Obershelp similarity on normalized text (30\%)
  {[}20, 21{]}, implemented via Python's
  \texttt{difflib.SequenceMatcher.ratio()}. The AST node type must match
  between deletion and addition candidates. Any pair similarity
  \(\geq 0.6\) (cost \(\leq 0.4\)) is classified as a ``modification''
  lineage rather than a true death. For large file-pairs
  (\(|\text{del}| \times |\text{add}| > 100{,}000\)), a greedy
  descending-similarity pass replaces the \(O(N^3)\) Hungarian step to
  bound memory usage.
\item
  \textbf{Cross-File Similarity (AST-aware):} Unmatched deletions and
  additions across different files are grouped by AST type and compared
  pairwise (groups exceeding 50,000 pairs are skipped). Handles
  directory renames and large-scale refactors not resolved by Stages
  1--3. Uses the same 0.6 threshold and Sørensen--Dice +
  Ratcliff/Obershelp composite.
\item
  \textbf{Intra-File AST-Agnostic Similarity (high-confidence only):} A
  final pass within each file path that removes the AST-type constraint
  but raises the threshold to \(\geq 0.9\). Resolves lines where the AST
  node type changed (e.g., an expression wrapped in a try-catch) but
  textual content is nearly identical.
\end{enumerate}

Conceptually aligning with the two-phase mapping approach of GumTree
{[}1{]}, our pipeline prioritizes exact structural AST matches (Stages
1--2). Stages 3--5 progressively relax constraints to capture
increasingly complex code evolutions while controlling for false
modification matches through higher thresholds and AST-type guards.

\hypertarget{structural-context-extraction}{%
\subsubsection{5.3 Structural Context
Extraction}\label{structural-context-extraction}}

Beyond the AST node type, we extract two categories of structural
features for each line by traversing the tree-sitter AST from the line's
node upward to the root:

\textbf{Nesting depth} is computed as the count of scope-creating
ancestors encountered during the parent-chain traversal. The following
AST node types increment the nesting counter: \texttt{statement\_block},
\texttt{class\_body}, \texttt{function\_declaration}, and
\texttt{arrow\_function}. Thus, a line inside a function body at the top
level has nesting depth 2 (one \texttt{function\_declaration} + one
\texttt{statement\_block}), while a line inside a nested callback has
depth 4 or higher. This metric captures the hierarchical complexity of a
line's position in the code structure.

\textbf{Four binary structural context flags} indicate whether any
ancestor in the parent chain matches specific scope types: -
\textbf{\texttt{in\_loop}}: Whether the line resides inside a
\texttt{for}, \texttt{for...in}, \texttt{while}, or \texttt{do} loop
body. - \textbf{\texttt{in\_condition}}: Whether the line is within the
body of an \texttt{if}, \texttt{switch}, or ternary expression. -
\textbf{\texttt{in\_try\_catch}}: Whether the line is enclosed in a
\texttt{try}, \texttt{catch}, or \texttt{finally} block. -
\textbf{\texttt{in\_function}}: Whether the line is inside a function
declaration, arrow function, or method definition.

\textbf{Line token count} (\texttt{log\_tokens}) quantifies the lexical
content of each line. We parse the line's source text with the
tree-sitter TypeScript grammar and count its \emph{leaf (terminal)
nodes} --- the lexer tokens of the line: identifiers, numeric/string
literals, keywords, operators, and punctuation. Multi-character
operators (\texttt{=\textgreater{}}, \texttt{===}) and string literals
are counted as the lexer emits them rather than split
character-by-character, so the count reflects lexical structure rather
than raw byte composition. For a line with \(T\) tokens we apply a
log-transform to stabilize variance and compress the long right tail:
\(\text{log\_tokens} = \ln(1 + T)\). High values correspond to lexically
rich lines (e.g., complex function calls, type annotations with multiple
generics); low values indicate short, simple statements. The count is
computed from each line's stored raw content using the same tree-sitter
parser used for AST typing (shared helper
\texttt{clsa\_libs.token\_utils}), so it is reproducible without version
history. (An earlier formulation of this feature used line Shannon
character \emph{entropy}; §6.5 reports why it was a line-size proxy and
replaced by the token count.)

\textbf{Line--TODO proximity} (\texttt{log\_todo\_distance}): For each
tracked line, we compute the absolute line-number distance to the
nearest \texttt{//\ TODO} or \texttt{//\ FIXME} annotation in the same
file at the time of the line's birth commit. If no such annotation
exists in the file, the distance is set to the file length. We then
apply a log-transform:
\texttt{log\_todo\_distance\ =\ ln(1\ +\ distance\_to\_todo)}, stored as
a \texttt{Float32} column in the \texttt{features} table. The feature
operationalizes proximity to acknowledged technical debt: a line
directly below a TODO marker is assigned a small value; a line in a
stable, annotation-free region of the file receives a large value. We
include it as a continuous covariate in the Cox model to test whether
proximity to known deferred-work annotations correlates with deletion
risk independently of structural features.

These features capture the enclosing structural scope, lexical content,
and proximity to deferred-work markers of a line. All binary and integer
features (nesting depth + four flags) are stored as \texttt{UInt8}
columns and \texttt{log\_todo\_distance} as a \texttt{Float32} column in
the ClickHouse \texttt{features} table, populated during the mining
phase; \texttt{log\_tokens} is derived from each line's stored raw
content at analysis time via the tree-sitter parser.

\hypertarget{storage-architecture}{%
\subsubsection{5.4 Storage Architecture}\label{storage-architecture}}

Processing results are stored in a ClickHouse columnar database using
\texttt{MergeTree} engines across nine domain tables: birth events,
death events, migration/modification lineage, per-line structural
features, raw content, and a cross-repository global dictionary for LIDF
computation. Full DDL and schema definitions are provided in the
replication package.

\hypertarget{statistical-analysis-design}{%
\subsubsection{5.5 Statistical Analysis
Design}\label{statistical-analysis-design}}

Our statistical pipeline consists of five complementary components:

\begin{enumerate}
\def\labelenumi{\arabic{enumi}.}
\tightlist
\item
  \textbf{Non-parametric estimation}: Kaplan--Meier survival curves with
  log-rank tests across stratified groups (RQ1--RQ4). All pairwise
  p-values are subject to Benjamini--Hochberg (BH) FDR correction. In
  the RQ2--RQ4 stratified log-rank analysis, 15 pairwise comparisons are
  performed: 10 RQ2 AST-group pairs (C(5,2) over the five most-populated
  AST groups --- \texttt{declaration}, \texttt{control\_flow},
  \texttt{expression}, \texttt{import\_export}, \texttt{type\_system};
  the heterogeneous \texttt{other} group is excluded from the
  KM/log-rank comparison but retained in the multivariate Cox model), 3
  RQ3 nesting-tier pairs (flat vs.~shallow, flat vs.~deep, shallow
  vs.~deep), and 2 RQ4 temporal pairs (weekday vs.~weekend, day
  vs.~night). 13 of 15 survive BH correction at α = 0.05; the two
  non-significant results are \texttt{control\_flow\ vs\ type\_system}
  (p\_adj = 0.988) and \texttt{declaration\ vs\ type\_system} (p\_adj =
  0.305).
\item
  \textbf{Semi-parametric modeling}: Cox Proportional Hazards regression
  with L2 regularization (penalizer = 0.1) and 15 covariates for
  multivariate assessment (RQ5). Proportional hazards assumption is
  explicitly tested via Schoenfeld residual diagnostics.
\item
  \textbf{Sensitivity analysis}: Four robustness checks---nested-only
  lines, single-repository, exclusion of ephemeral lines (\textless1
  day), and a per-repository stratified subsample (cap 10,000 lines per
  repository)---plus Weibull AFT and Log-Logistic AFT models to evaluate
  covariate stability under distributional alternatives.
\item
  \textbf{Time-stratified Cox analysis}: Separate Cox models on three
  landmark time bands (0--90d, 90--365d, 365+d) to expose time-varying
  covariate effects and provide mechanistic interpretation of PH
  violations.
\item
  \textbf{Shared frailty model}: A gamma frailty Cox model with
  repository identity as cluster variable
  (\(Z_i \sim \mathrm{Gamma}(1/\theta, 1/\theta)\)), fitted in R
  (\texttt{survival} package) to decompose within-project and
  between-project sources of hazard variation and address Simpson's
  paradox observed in the marginal model.
\end{enumerate}

\textbf{Sample sizes across analyses.} Different analytical stages use
different sample sizes, each motivated by a specific constraint: the
Python Cox model uses n = 300,000 (subsample of 350K, drawn to manage
Schoenfeld residual computation cost); the time-stratified landmark
analysis uses the full sample (349,510 lines after dropping
comment\_like; no Schoenfeld computation required); and both R models
(marginal and frailty Cox) use n = 346,808 (the full 350,000-line
analytical sample after dropping comment\_like lines and excluding
repositories with fewer than 30 events from frailty clustering).
Numerical differences in hazard ratios between Python and R models are
additionally attributable to the L2 ridge penalty applied only in the
Python model (λ = 0.1), which shrinks coefficients toward zero.

\begin{center}\rule{0.5\linewidth}{0.5pt}\end{center}

\hypertarget{empirical-evaluation}{%
\subsection{6. Empirical Evaluation}\label{empirical-evaluation}}

\hypertarget{dataset-summary}{%
\subsubsection{6.1 Dataset Summary}\label{dataset-summary}}

We mined 120 active TypeScript repositories spanning a wide range of
project ages, sizes, and domains. In total, the dataset contains 32.5
million line-level birth events across all repositories, of which 11.0
million (33.9\%) are true deletions (\texttt{hard\_delete} +
\texttt{file\_delete}); the remaining 21.5 million (66.1\%) are
right-censored --- lines that remain alive or underwent
migration/modification as of the repository's last commit.

For the Cox analytical design, we sampled 350,000 lines using a
deterministic hash-based sampling strategy
(\texttt{ORDER\ BY\ cityHash64(id)\ LIMIT\ 350000}), ensuring full
reproducibility. After censoring detection and quality filtering, the
final dataset contains:

\begin{itemize}
\tightlist
\item
  \textbf{350,000 total observations}
\item
  \textbf{118,787 death events} (33.9\%)
\item
  \textbf{231,213 right-censored observations} (66.1\%)
\end{itemize}

Table 1 presents the top 10 repositories by total lines mined. The full
catalog of all 120 repositories is provided in Appendix A.

\begin{longtable}[]{@{}
  >{\raggedright\arraybackslash}p{(\columnwidth - 12\tabcolsep) * \real{0.2267}}
  >{\raggedright\arraybackslash}p{(\columnwidth - 12\tabcolsep) * \real{0.1467}}
  >{\raggedright\arraybackslash}p{(\columnwidth - 12\tabcolsep) * \real{0.1200}}
  >{\raggedright\arraybackslash}p{(\columnwidth - 12\tabcolsep) * \real{0.0933}}
  >{\raggedright\arraybackslash}p{(\columnwidth - 12\tabcolsep) * \real{0.1733}}
  >{\raggedright\arraybackslash}p{(\columnwidth - 12\tabcolsep) * \real{0.1067}}
  >{\raggedright\arraybackslash}p{(\columnwidth - 12\tabcolsep) * \real{0.1333}}@{}}
\toprule\noalign{}
\begin{minipage}[b]{\linewidth}\raggedright
Repository Name
\end{minipage} & \begin{minipage}[b]{\linewidth}\raggedright
Age (Mo.)
\end{minipage} & \begin{minipage}[b]{\linewidth}\raggedright
Commits
\end{minipage} & \begin{minipage}[b]{\linewidth}\raggedright
Files
\end{minipage} & \begin{minipage}[b]{\linewidth}\raggedright
Lines Mined
\end{minipage} & \begin{minipage}[b]{\linewidth}\raggedright
Deaths
\end{minipage} & \begin{minipage}[b]{\linewidth}\raggedright
Survived
\end{minipage} \\
\midrule\noalign{}
\endhead
\bottomrule\noalign{}
\endlastfoot
microsoft/\discretionary{}{}{}vscode & 126 & 118,215 & 18,375 &
4,031,373 & 966,838 & 1,909,391 \\
Expensify/App & 35 & 99,575 & 12,513 & 2,835,458 & 1,039,257 &
1,124,026 \\
twentyhq/\discretionary{}{}{}twenty & 41 & 10,201 & 44,565 & 2,170,359 &
785,557 & 990,896 \\
DefinitelyTyped/\discretionary{}{}{}DefinitelyTyped & 163 & 69,085 &
19,547 & 1,853,038 & 724,178 & 773,375 \\
ag-grid/\discretionary{}{}{}ag-grid & 130 & 23,141 & 20,475 & 1,570,216
& 539,199 & 693,573 \\
microsoft/\discretionary{}{}{}FluidFramework & 116 & 16,807 & 17,976 &
1,468,504 & 558,400 & 569,540 \\
remotion-dev/\discretionary{}{}{}remotion & 71 & 18,733 & 10,709 &
1,349,990 & 328,921 & 614,026 \\
tamagui/\discretionary{}{}{}tamagui & 67 & 11,132 & 15,035 & 1,132,232 &
688,627 & 285,154 \\
supabase/\discretionary{}{}{}supabase & 67 & 15,808 & 13,485 & 1,072,324
& 355,801 & 541,225 \\
triggerdotdev/\discretionary{}{}{}trigger.\discretionary{}{}{}dev & 41 &
4,979 & 4,552 & 1,019,397 & 629,277 & 329,335 \\
\end{longtable}

\emph{Table 1: Dataset Overview (top 10 by total lines mined; 120
repositories total). Note: }Deaths* + \emph{Survived} \textless{}
\emph{Lines Mined} in all rows; the gap represents lines whose identity
was transformed via migration or modification (logged in
\texttt{line\_evolution}) --- these are right-censored in the survival
model but do not appear in either the death or survived counts. The
complete repository catalog is provided in Appendix A.*

\hypertarget{data-cleaning-construct-validity}{%
\subsubsection{6.2 Data Cleaning \& Construct
Validity}\label{data-cleaning-construct-validity}}

The pipeline operates on two parallel streams. On the \textbf{birth
(additions) side}, the AST-aware filter removed \textbf{4,807,296}
comment lines identified via tree-sitter semantics and
\textbf{19,626,602} trivial structural tokens (length \textless{} 5)
from the raw Git additions, yielding \textbf{32,464,566} tracked birth
events across all 120 repositories. On the \textbf{death (deletions)
side}, the multi-stage bipartite alignment prevented \textbf{8,293,504}
false deaths by resolving them as semantic migrations or modifications
(logged in the \texttt{line\_evolution} table), yielding a highly
purified dataset of \textbf{11,009,579} actual line-level extinctions.

\hypertarget{baseline-survival-rq1}{%
\subsubsection{6.3 Baseline Survival
(RQ1)}\label{baseline-survival-rq1}}

Our baseline Kaplan--Meier estimator crosses the 50\% survival threshold
within the observation window. The Kaplan--Meier median survival time is
\textbf{1,335 days} (about 3.7 years) on the 350,000-line analytical
sample; we checked this two further ways to rule out a sampling artifact
--- on the full 32.5-million-line population (median 1,352 days) and on
an independent 350,000-line resample drawn with a different hash
function (median 1,339 days) --- all three estimates agree within three
weeks of each other. Among lines that are actually deleted, the median
time-to-deletion is \textbf{95.7 days}.

This heavy-tailed distribution is consistent with a ``stabilize or die''
pattern: lines that survive the initial refactoring phases have a
substantially reduced deletion hazard, even though most are eventually
deleted given enough time. This directionally corroborates Spinellis et
al.~{[}25{]}, whose Weibull shape parameter below one in every project
indicates the same \emph{infant-mortality} regime --- deletion hazard is
highest for young lines and decreases with age --- which our
time-stratified analysis (§6.9) makes explicit. The empirical threshold
of this early-fragile period corresponds to the median lifespan of
deleted lines --- \textbf{95.7 days} --- so code surviving past
\textasciitilde100 days is already outlasting the typical deleted line.
By the end of the observation window (repositories span 6--163 months),
the Kaplan--Meier estimate has fallen to 38.6\% survival; the median
lies well inside the window, at roughly one-third of the longest
repository history (163 months ≈ 4,960 days), so the crossing is a
property of the corrected data rather than an artifact of an unusually
short window. An earlier version of this analysis, using an uncorrected
censoring treatment that credited rewritten lines with survival time
through the repository's last commit rather than through the commit that
rewrote them, found the curve never crossing 50\% within the window;
correcting that treatment moves roughly a fifth of the sample's
censoring times substantially earlier and is sufficient on its own to
produce the crossing reported here.

\begin{figure}
\centering
\includegraphics{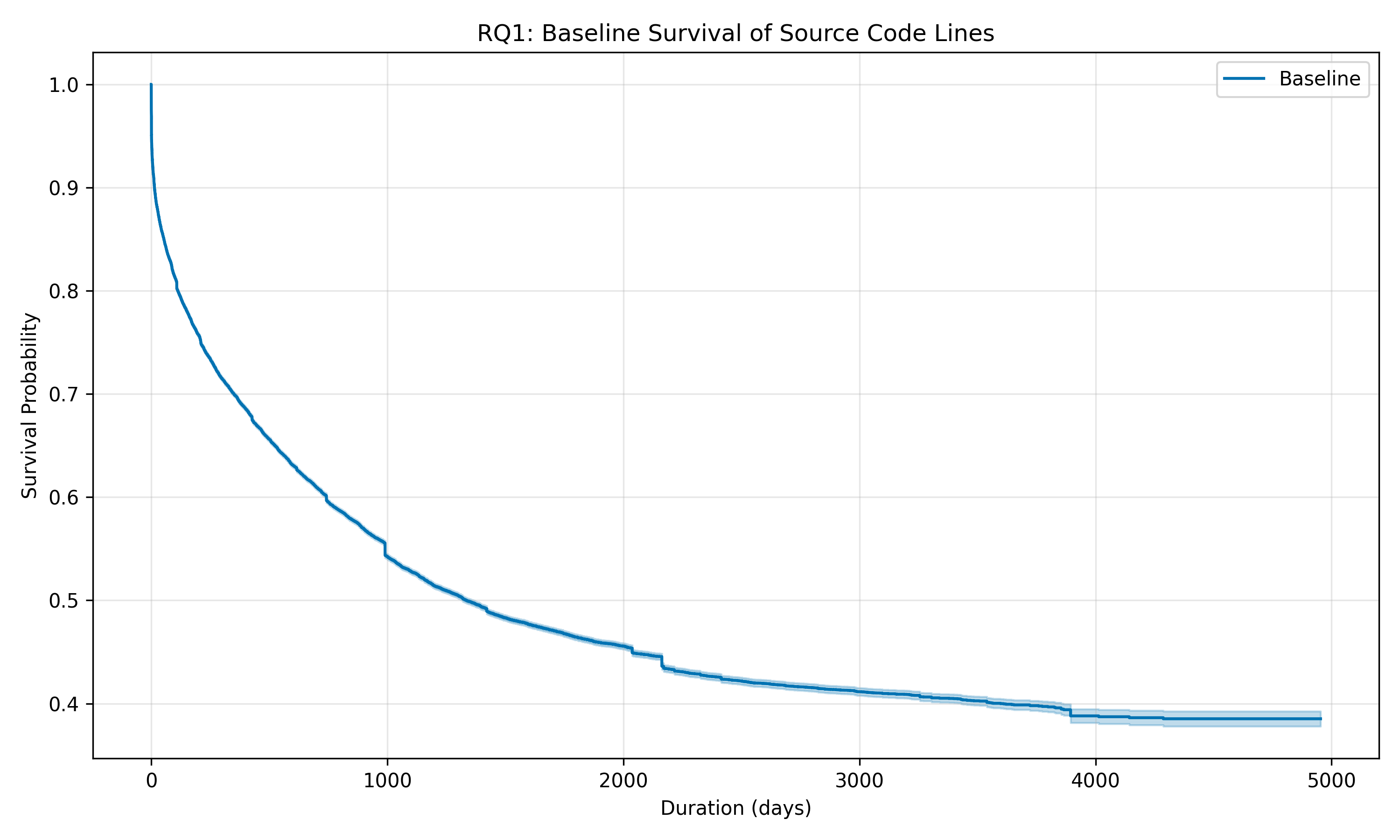}
\caption{Baseline Kaplan--Meier survival curve for all source code lines
(solid line; shaded band shows the 95\% confidence interval). The
survival function crosses 0.50 at approximately 1,335 days and reaches
38.6\% by the end of the observation window.}
\end{figure}

\hypertarget{stratified-group-comparisons-rq2-rq3-rq4}{%
\subsubsection{6.4 Stratified Group Comparisons (RQ2, RQ3,
RQ4)}\label{stratified-group-comparisons-rq2-rq3-rq4}}

We conducted stratified log-rank tests to differentiate hazard dynamics
across structural and temporal dimensions. All reported p-values are
corrected using Benjamini--Hochberg FDR. For the RQ2--RQ4 stratified
log-rank tests, 15 pairwise comparisons are performed (10 RQ2 AST-group
pairs over the five most-populated AST groups, C(5,2), excluding the
heterogeneous \texttt{other} group from the KM comparison; 3 RQ3
nesting-tier pairs; 2 RQ4 temporal comparisons); 13 of 15 survive
correction at α = 0.05 (see §5.5 for the complete breakdown).

\begin{itemize}
\item
  \textbf{RQ2 (Syntax Role):} Pairwise log-rank tests across the five
  most-populated AST groups confirm that syntactic role is strongly
  associated with survival: 8 of the 10 group pairs are significant
  after BH correction; the two exceptions both involve
  \texttt{type\_system}, the smallest of the five groups ---
  \texttt{control\_flow\ vs\ type\_system} (χ² ≈ 0.0, p\_adj = 0.988)
  and \texttt{declaration\ vs\ type\_system} (χ² = 1.1, p\_adj = 0.305)
  --- while \texttt{type\_system} still separates cleanly from the two
  largest, most volatile groups (\texttt{expression},
  \texttt{import\_export}). The widest separation is
  \texttt{expression\ vs\ import\_export} (χ² = 636, p\_adj ≈ 10⁻¹³⁹).
  For per-group effect sizes we report the multivariate Cox hazard
  ratios (full model in §6.5, Table 2), relative to the reference
  category (control flow); the five non-reference groups vary
  substantially in magnitude and several exhibit instability under
  sensitivity analysis:

  \begin{itemize}
  \tightlist
  \item
    \textbf{Import/export} statements (HR = 0.74, 95\% CI {[}0.72,
    0.76{]}): The strongest protective effect among AST groups. Import
    lines are structural scaffolding---rarely modified once established.
    However, this effect is \textbf{unstable}: in the nested-only
    submodel, the estimate becomes unreliable (HR = 0.19, 95\% CI
    {[}0.01, 2.79{]}, ns) because nested imports are rare, yielding very
    few observations in that stratum.
  \item
    \textbf{Declaration} lines (HR = 0.84, 95\% CI {[}0.81, 0.88{]}): A
    substantial protective effect, consistent with declarations
    (variable/function/class signatures) serving as anchor points. This
    effect is \textbf{unstable in sensitivity analysis}: it becomes
    non-significant in both the nested-only (HR = 0.95, ns) and
    single-repo (HR = 0.95, ns) models, suggesting partial confounding
    with repository-level coding conventions.
  \item
    \textbf{Expression} lines (HR = 1.23, 95\% CI {[}1.19, 1.27{]}): The
    strongest risk factor in the model---the highest hazard among all
    covariates. Expression statements (function calls, assignments) are
    the ``working'' lines of code, subject to frequent modification as
    logic evolves. This effect is \textbf{stable and consistent across
    all sensitivity models}.
  \item
    \textbf{Other} (HR = 1.16, 95\% CI {[}1.13, 1.18{]}): A moderate
    risk factor. Miscellaneous AST nodes (JSX, template literals, etc.)
    are notably more volatile than the control-flow baseline.
  \item
    \textbf{Type system} constructs (HR = 0.82, 95\% CI {[}0.74,
    0.91{]}): Relative to control flow, this effect is significant
    (\emph{p} \textless{} 0.001) and directionally consistent across all
    sensitivity models and both AFT models, though it loses significance
    in the single-repo model (HR = 0.66, ns) due to the small number of
    nested type-system lines within a single project.
  \end{itemize}

  The reference category is \textbf{control flow} (\texttt{if},
  \texttt{switch}, \texttt{for}, \texttt{while}, ternary constructs)---a
  well-populated, interpretable structural group. We do not use
  \texttt{block}-level tokens as the baseline: statement-block and
  class-body lines are almost entirely short structural delimiters
  (\texttt{\{}, \texttt{\}}) removed by the trivial-token filter (§5.1),
  so that group is effectively empty in the modelled data. The same
  \texttt{control\_flow} reference is used in the R frailty model (Table
  7), making the marginal and frailty AST hazard ratios directly
  comparable.

  \begin{figure}
  \centering
  \includegraphics{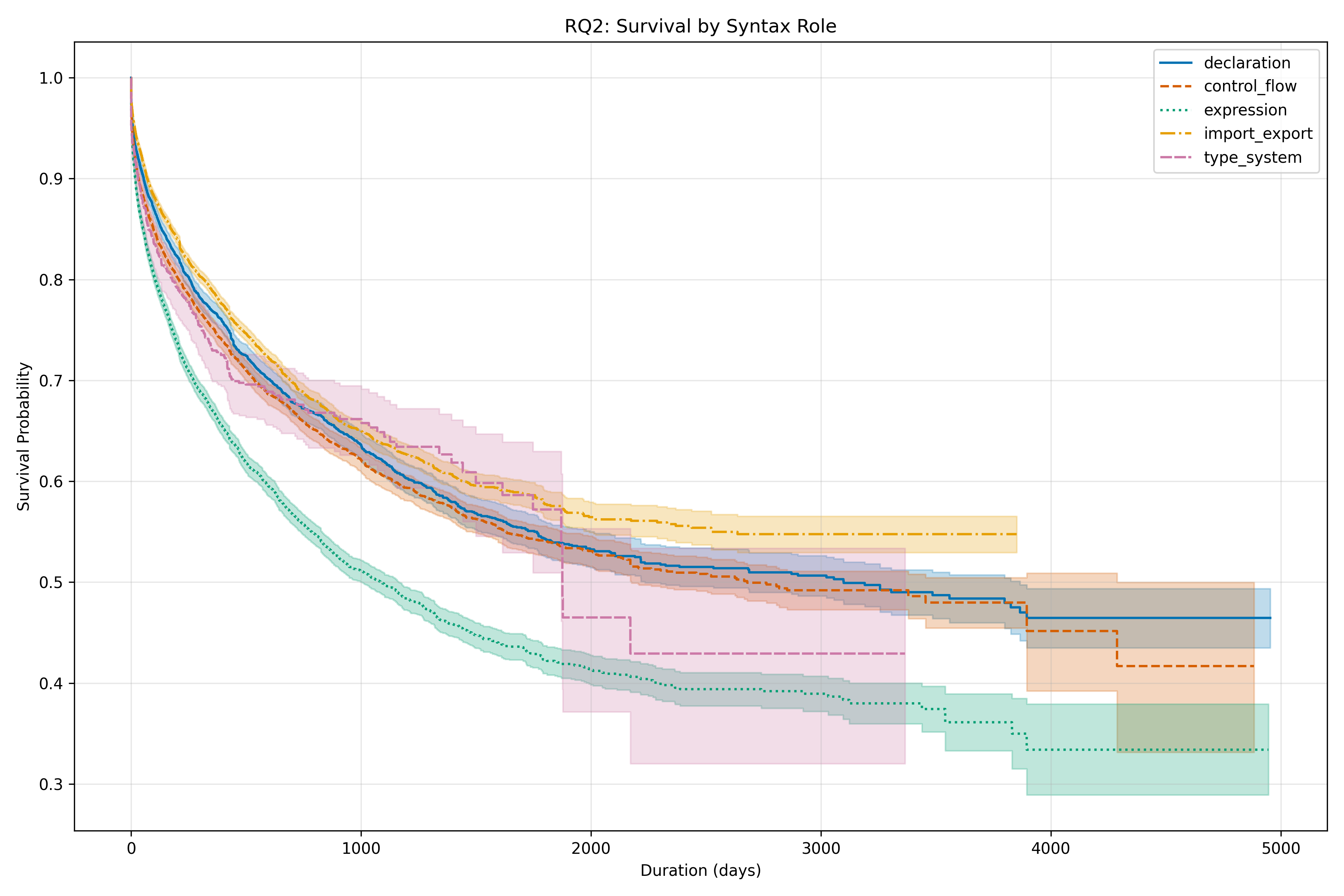}
  \caption{Kaplan--Meier survival curves stratified by AST group ---
  declaration, control flow, expression, import/export, and type system
  (shaded bands are 95\% confidence intervals). Expression lines show
  the steepest decline and the lowest survivorship, whereas
  import/export lines retain the highest survival, consistent with the
  hazard ratios in Table 2.}
  \end{figure}
\item
  \textbf{RQ3 (Structural Complexity):} Lines are partitioned into three
  \textbf{mutually exclusive} nesting tiers --- flat (L = 0; n =
  144,164), shallow (L = 1--2; n = 112,221), and deep (L \textgreater{}
  2; n = 93,615) --- so that each pairwise log-rank test contrasts
  disjoint populations. All three comparisons are significant after BH
  correction: flat vs.~shallow (χ² = 1571), flat vs.~deep (χ² = 2728),
  and shallow vs.~deep (χ² = 228, p\_adj \textless{} 10⁻⁵⁰).
  Survivorship increases monotonically with nesting depth: the censoring
  rate rises from 62.0\% (flat) to 66.8\% (shallow) to 71.5\% (deep),
  the Kaplan--Meier median is reached only for flat lines (2,066 days)
  and not within the observation window for either nested tier, and
  among lines that do die the median lifespan is 57.1 days (flat), 116.0
  days (shallow) and 118.9 days (deep). Flat lines therefore survive
  markedly less than nested ones, while the shallow--deep gap is real
  but small. In the multivariate Cox model, deeper nesting modestly
  decreases hazard (HR = 0.97 per nesting level, 95\% CI {[}0.96,
  0.97{]}), and the direction is \textbf{consistent across all
  sensitivity models} (see §6.7).

  The novel structural context covariates provide additional granularity
  beyond nesting depth:

  \begin{itemize}
  \tightlist
  \item
    \textbf{\texttt{in\_function}} (HR = 0.81, 95\% CI {[}0.80,
    0.82{]}): Lines inside function bodies exhibit a strong and
    \textbf{stable} protective effect, consistent across all sensitivity
    models (HR range: 0.80--0.95). This is among the strongest
    structural predictors in the model.
  \item
    \textbf{\texttt{in\_condition}} (HR = 1.13, 95\% CI {[}1.12,
    1.15{]}): Lines within conditional branches face elevated hazard,
    consistent across all models (see §6.7).
  \item
    \textbf{\texttt{in\_loop}} (HR = 0.96, 95\% CI {[}0.93, 0.99{]},
    \emph{p} = 0.02): Lines inside loop bodies show a modest protective
    effect in the aggregate model. \textbf{Note:} this effect loses
    significance under within-commit clustered sandwich SE correction
    (§8.2) and disappears entirely after frailty conditioning (§6.10);
    it should be treated as a repository-level proxy rather than a
    genuine line-level predictor.
  \item
    \textbf{\texttt{in\_try\_catch}} (HR = 1.07, 95\% CI {[}1.03,
    1.12{]}, \emph{p} \textless{} 0.005): Lines within error-handling
    blocks face modestly elevated hazard in the aggregate model, though
    this effect loses significance after frailty conditioning (§6.10).
  \end{itemize}

  \begin{figure}
  \centering
  \includegraphics{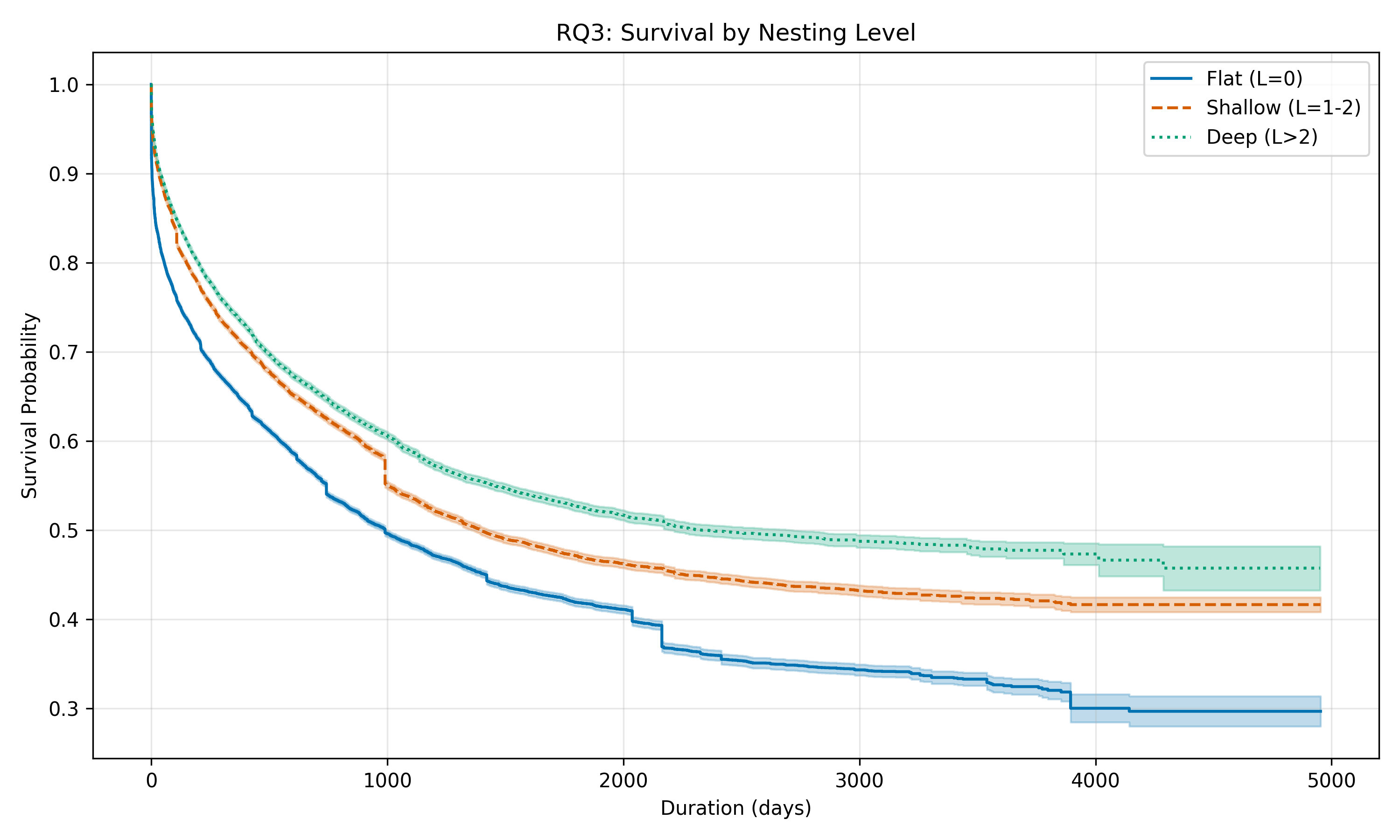}
  \caption{Kaplan--Meier survival curves by mutually exclusive nesting
  tier: flat (L = 0), shallow (L = 1--2) and deep (L \textgreater{} 2).
  The three curves separate cleanly and are ordered monotonically in
  depth. Flat lines die fastest; the flat-to-shallow gap is far larger
  than the shallow-to-deep gap, so the protective effect of nesting is
  concentrated at the first scope boundary and continues only weakly
  with further depth. The effect is partly repository-dependent (see
  §6.7 and §6.10).}
  \end{figure}
\item
  \textbf{RQ4 (Temporal Context):} \textbf{Nocturnal commits}
  (22:00--05:00) are associated with elevated hazard in the aggregate
  model (HR = 1.16, 95\% CI {[}1.15, 1.18{]}, \emph{p} \textless{}
  0.005). \textbf{Note:} \texttt{is\_night} attenuates substantially
  under repository-level conditioning --- it stays significant in the
  single-repository sensitivity model (HR = 1.06, §6.7) but is reduced
  to HR = 1.13 under the shared frailty model (§6.10) --- indicating
  that part, but not all, of this association is driven by
  between-repository confounding. \textbf{Weekend} commits are not
  significant in the aggregate model (HR = 1.00, 95\% CI {[}0.99,
  1.02{]}, \emph{p} = 0.67) and remain ns in the stratified model, but
  become a significant risk factor in both the single-repository (HR =
  1.11) and no-ephemeral (HR = 1.03) models --- a repository- and
  duration-level confound discussed in §6.7. The day-versus-night
  log-rank comparison is highly significant (χ² = 715, p\_adj ≈ 10⁻¹⁵⁷).
  The weekday-versus-weekend comparison is significant at this sample
  size (χ² = 5.6, p\_adj = 0.023) but the effect is null after adjusting
  for other covariates (Table 2, \texttt{is\_weekend} HR ≈ 1.00, ns) ---
  a difference detectable in the raw group comparison that a
  multivariate model correctly assigns no independent weight.

  We emphasize that this is an \emph{association}, not a causal claim.
  Nocturnal commits may correlate with urgency-driven fixes,
  experimental feature branches, or timezone artifacts.

  \begin{figure}
  \centering
  \includegraphics{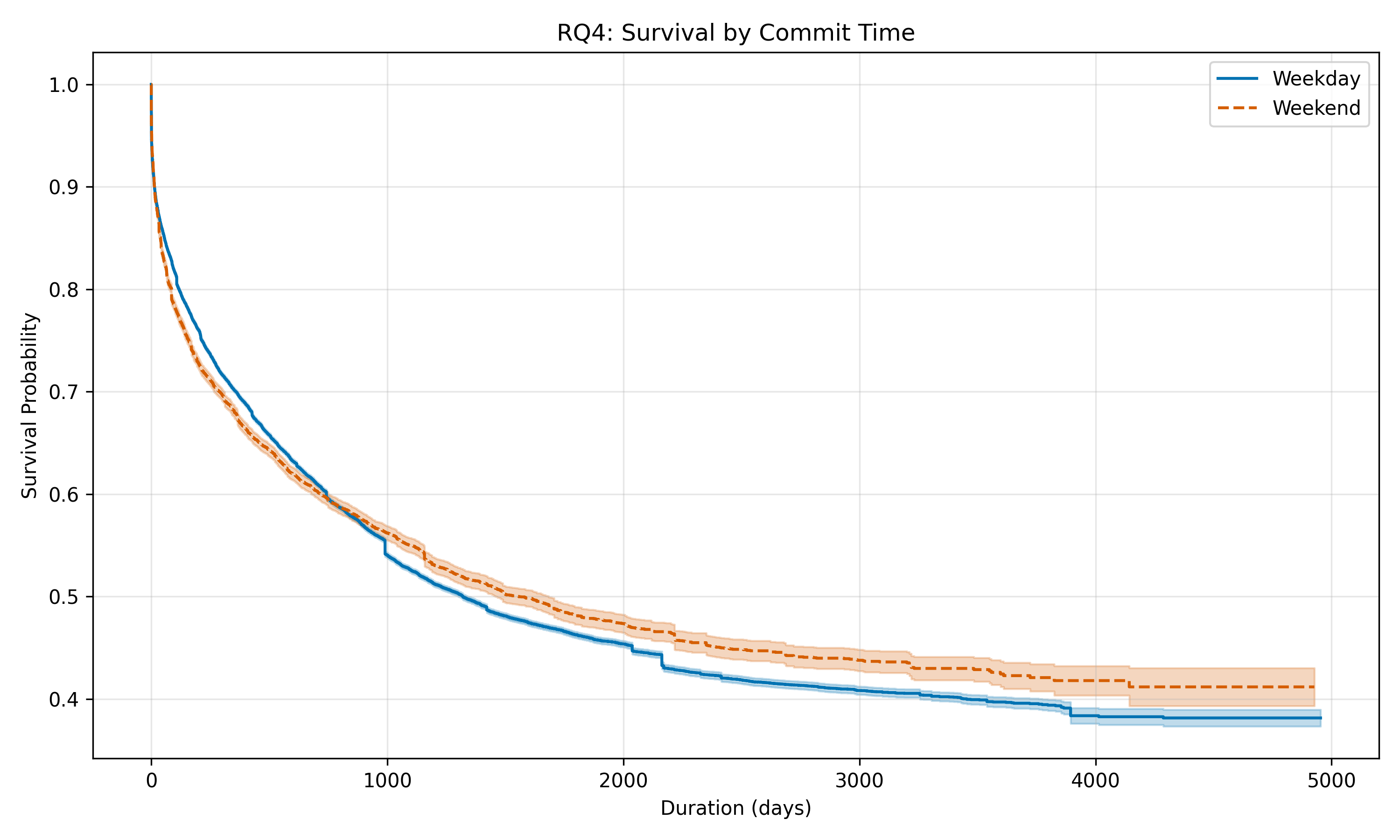}
  \caption{Kaplan--Meier survival curves stratified by commit day type
  --- weekday vs.~weekend (shaded bands are 95\% confidence intervals).
  The two curves are close but, at this sample size, statistically
  distinguishable (§6.4); the Cox model's \texttt{is\_weekend} covariate
  (Table 2, HR ≈ 1.00) shows no residual effect once other covariates
  are held fixed. The day-vs-night contrast emphasized in the text is
  captured by the \texttt{is\_night} covariate in the Cox models (Table
  2), not by this plot.}
  \end{figure}
\end{itemize}

\hypertarget{multivariate-cox-proportional-hazards-model-rq5}{%
\subsubsection{6.5 Multivariate Cox Proportional Hazards Model
(RQ5)}\label{multivariate-cox-proportional-hazards-model-rq5}}

The 15-covariate Cox PH model identifies \textbf{import/export lines} as
the strongest protective factor (HR = 0.74) and \textbf{expression-type
AST nodes} as the strongest risk predictor (HR = 1.23), with
\textbf{\texttt{in\_function}} (HR = 0.81) and \textbf{line token count}
(HR = 0.87) close behind on the protective side, fitted on 300,000 lines
with 101,773 deletion events (subsample drawn uniformly at random,
\texttt{random\_state=42}; full summary in Table 2).

\emph{A note on the lexical feature.} An earlier formulation used line
Shannon character entropy (\texttt{log\_entropy}) in place of the token
count. Following a reviewer observation (D. M. Jones, pers. comm.), we
verified that character entropy is, by the identity
\(H = \log_2 L - (1/L)\sum_c n_c \log_2 n_c\), dominated by the
logarithm of line length \(L\): regressing \(H\) on \(\log_2 L\) gives
\(R^2 = 0.75\). In standardized (per-SD) Cox models, entropy and
character length converge to the same near-null hazard ratio (HR
\(\approx\) 0.976 each), and entropy's length-orthogonal residual,
though still statistically significant at this sample size, is
practically negligible (HR \(\approx\) 0.99 per SD). Token count is
\emph{not} reducible to character length --- it retains a
size-independent survival signal (HR \(\approx\) 0.90 per SD) and yields
the best standalone concordance (C = 0.587, vs \$\approx\$0.583 for both
character length and entropy). We therefore adopt \texttt{log\_tokens}
as the lexical covariate throughout; the full confound analysis is
provided in the supplementary material.

\begin{longtable}[]{@{}lllll@{}}
\toprule\noalign{}
Covariate & HR & 95\% CI & Effect & Sig. \\
\midrule\noalign{}
\endhead
\bottomrule\noalign{}
\endlastfoot
\texttt{ast\_\discretionary{}{}{}group\_\discretionary{}{}{}expression}
& 1.23 & {[}1.19, 1.27{]} & ↑risk 23.1\% & *** \\
\texttt{is\_\discretionary{}{}{}night} & 1.16 & {[}1.15, 1.18{]} & ↑risk
16.3\% & *** \\
\texttt{ast\_\discretionary{}{}{}group\_\discretionary{}{}{}other} &
1.16 & {[}1.13, 1.18{]} & ↑risk 15.7\% & *** \\
\texttt{in\_\discretionary{}{}{}condition} & 1.13 & {[}1.12, 1.15{]} &
↑risk 13.4\% & *** \\
\texttt{in\_\discretionary{}{}{}try\_\discretionary{}{}{}catch} & 1.08 &
{[}1.04, 1.13{]} & ↑risk 8.1\% & *** \\
\texttt{log\_\discretionary{}{}{}todo\_\discretionary{}{}{}distance} &
1.02 & {[}1.01, 1.02{]} & ↑risk 1.5\% & *** \\
\texttt{lidf} & 1.01 & {[}1.01, 1.02{]} & ↑risk 1.3\% & *** \\
\texttt{is\_\discretionary{}{}{}weekend} & 1.00 & {[}0.99, 1.02{]} &
↑risk 0.4\% & ns \\
\texttt{nesting\_\discretionary{}{}{}level} & 0.96 & {[}0.96, 0.97{]} &
↓risk 3.7\% & *** \\
\texttt{in\_\discretionary{}{}{}loop} & 0.96 & {[}0.93, 0.99{]} & ↓risk
4.1\% & * \\
\texttt{log\_\discretionary{}{}{}tokens} & 0.87 & {[}0.86, 0.88{]} &
↓risk 12.8\% & *** \\
\texttt{ast\_\discretionary{}{}{}group\_\discretionary{}{}{}declaration}
& 0.84 & {[}0.81, 0.88{]} & ↓risk 15.7\% & *** \\
\texttt{ast\_\discretionary{}{}{}group\_\discretionary{}{}{}type\_\discretionary{}{}{}system}
& 0.82 & {[}0.74, 0.91{]} & ↓risk 18.0\% & *** \\
\texttt{in\_\discretionary{}{}{}function} & 0.81 & {[}0.80, 0.82{]} &
↓risk 19.0\% & *** \\
\texttt{ast\_\discretionary{}{}{}group\_\discretionary{}{}{}import\_\discretionary{}{}{}export}
& 0.74 & {[}0.72, 0.76{]} & ↓risk 26.1\% & *** \\
\end{longtable}

\emph{Table 2: Cox PH results (main model, 15 covariates). *** denotes p
\textless{} 0.001, ** denotes p \textless{} 0.01, * denotes p
\textless{} 0.05, ns = not significant. Reference category:
\texttt{ast\_group} = \texttt{control\_flow} (see §6.4). The
\texttt{ast\_group\_control\_flow} dummy is therefore absorbed into the
baseline, and comment/shebang lines (\texttt{comment\_like}) are
excluded; the model has 15 covariates.}

\textbf{Concordance (C-index) = 0.586.} Most covariates are
statistically significant in the main model (14 of 15; only
\texttt{is\_weekend} is not), but the model's discriminative power is
limited. A C-index of 0.586 falls between random (0.50) and practically
useful (\textgreater0.65). The pattern reflects high population
heterogeneity---structural features explain \emph{some} variance in code
lifespan but are far from deterministic. We caution that statistical
significance should not be conflated with practical significance for
covariates with HR close to 1.0, particularly \texttt{lidf} (HR = 1.01,
statistically significant at this sample size but a small effect) and
\texttt{in\_loop} (HR = 0.96) --- both are near-null and unstable under
frailty conditioning (§6.10), indicating a partial repository-level
component alongside any genuine line-level effect.
\texttt{log\_todo\_distance} (HR = 1.02), though modest, is confirmed as
a genuine within-project effect in §6.10. \texttt{is\_weekend} is not
significant in the main model (HR = 1.00, p = 0.67) and remains
unreliable across most sensitivity models --- it should not be treated
as a reliable predictor.

\begin{figure}
\centering
\includegraphics{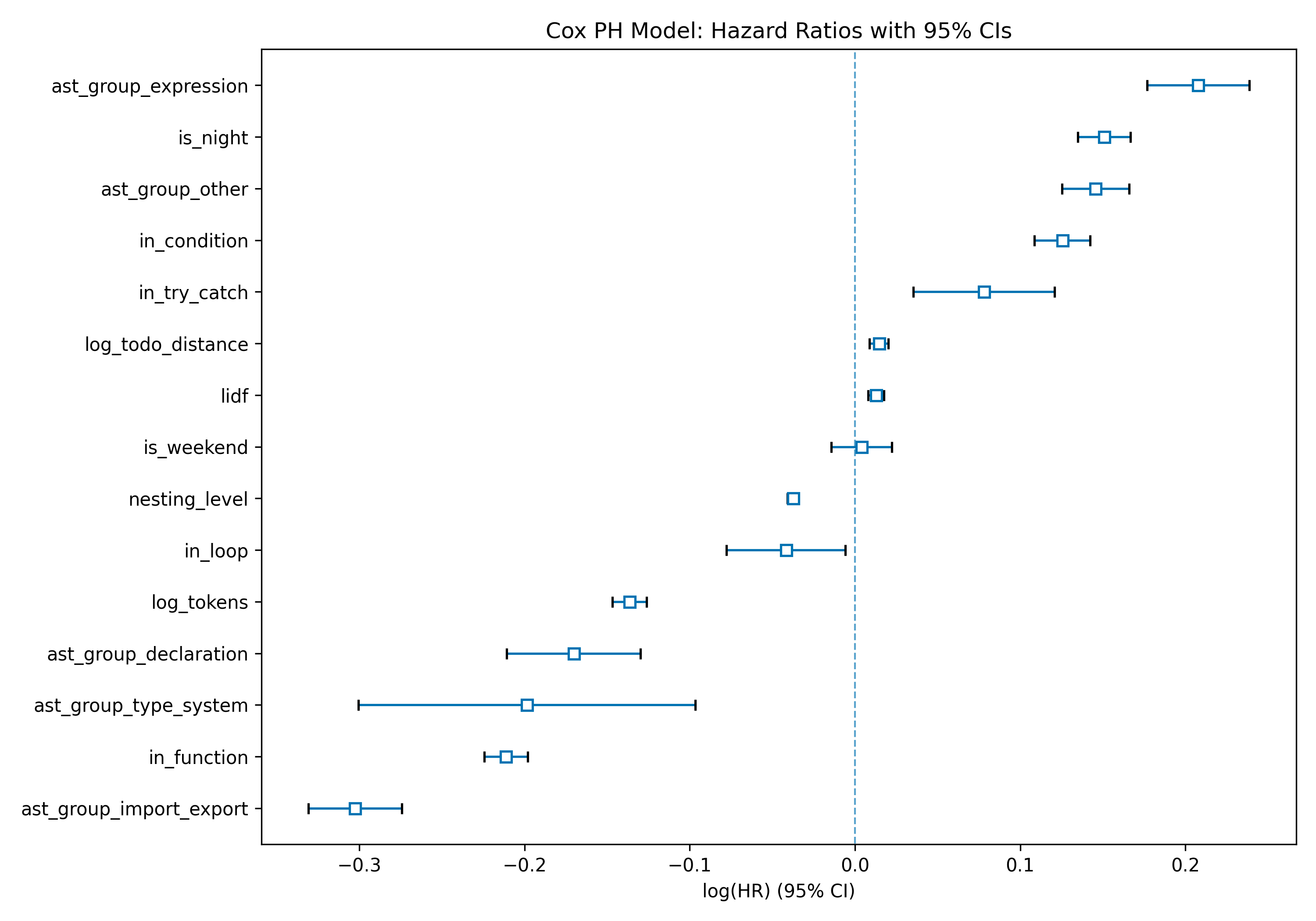}
\caption{Forest plot of log(HR) with 95\% confidence intervals for all
15 covariates of the main Cox model (vertical dashed line marks the null
at log(HR) = 0; markers to the right indicate elevated hazard, to the
left reduced hazard). \texttt{ast\_group\_expression} is the strongest
risk factor, while \texttt{ast\_group\_import\_export}, line token count
(\texttt{log\_tokens}), and \texttt{in\_function} are the strongest
protective factors.}
\end{figure}

\textbf{Key findings}: 1. \textbf{Expression-type lines} are the
dominant risk factor (HR = 1.23, ↑23.1\%). Lines containing function
calls and assignments are the most volatile AST group, with an effect
that is \textbf{stable across all sensitivity models}. Nocturnal commits
are also a significant risk factor (HR = 1.16, ↑16.3\%), now slightly
exceeding \texttt{ast\_group\_other} (HR = 1.16, ↑15.7\%) in magnitude.
The \texttt{is\_night} effect is consistent across all four sensitivity
models (see §6.7). 2. \textbf{Line token count} is a strong protective
lexical factor (HR = 0.87, ↓12.8\%): lexically richer lines are less
likely to die at any given time than short, simple ones. The strongest
protective factors overall are import/export lines (HR = 0.74) and
\texttt{in\_function} (HR = 0.81). The token-count effect is
\textbf{robust across all sensitivity checks} (see §6.7). 3.
\textbf{\texttt{in\_function}} shows a strong and stable protective
effect (HR = 0.81, ↓19.0\%), consistent across all sensitivity models.
\textbf{LIDF} (Line Inverse Document Frequency) shows a small but
statistically significant risk effect (HR = 1.01, ↑1.3\%), weaker than
previously estimated; whether this reflects a genuine within-project
signal or residual dependence on line size is not resolved by this model
alone (see §6.7 for stability across sensitivity checks). 4.
\textbf{\texttt{log\_todo\_distance}} is a modest but significant risk
factor (HR = 1.02, ↑1.5\%): lines further from TODO comments face
slightly higher hazard. This effect is confirmed as a genuine
within-project signal under frailty conditioning (see §6.10), where it
remains significant in both marginal (HR = 1.016) and frailty (HR =
1.024) R models.

\hypertarget{change-type-awareness-rq6}{%
\subsubsection{6.6 Change-Type Awareness
(RQ6)}\label{change-type-awareness-rq6}}

The three-way change classification (migration, modification, true
death) described in §5.2 prevents 8.3 million false deaths from
inflating hazard estimates. KM curves stratified by first change type
(Figure 6) confirm that migrations occur earliest, followed by
modifications, while true hard-deletion events exhibit the heaviest
right tail. Migrations and modifications are treated as right-censored
in all multivariate models. A formal competing risks analysis
(Fine--Gray sub-distribution hazard) is deferred to future work;
potential informative censoring from this treatment is discussed in
§8.1.

\begin{figure}
\centering
\includegraphics{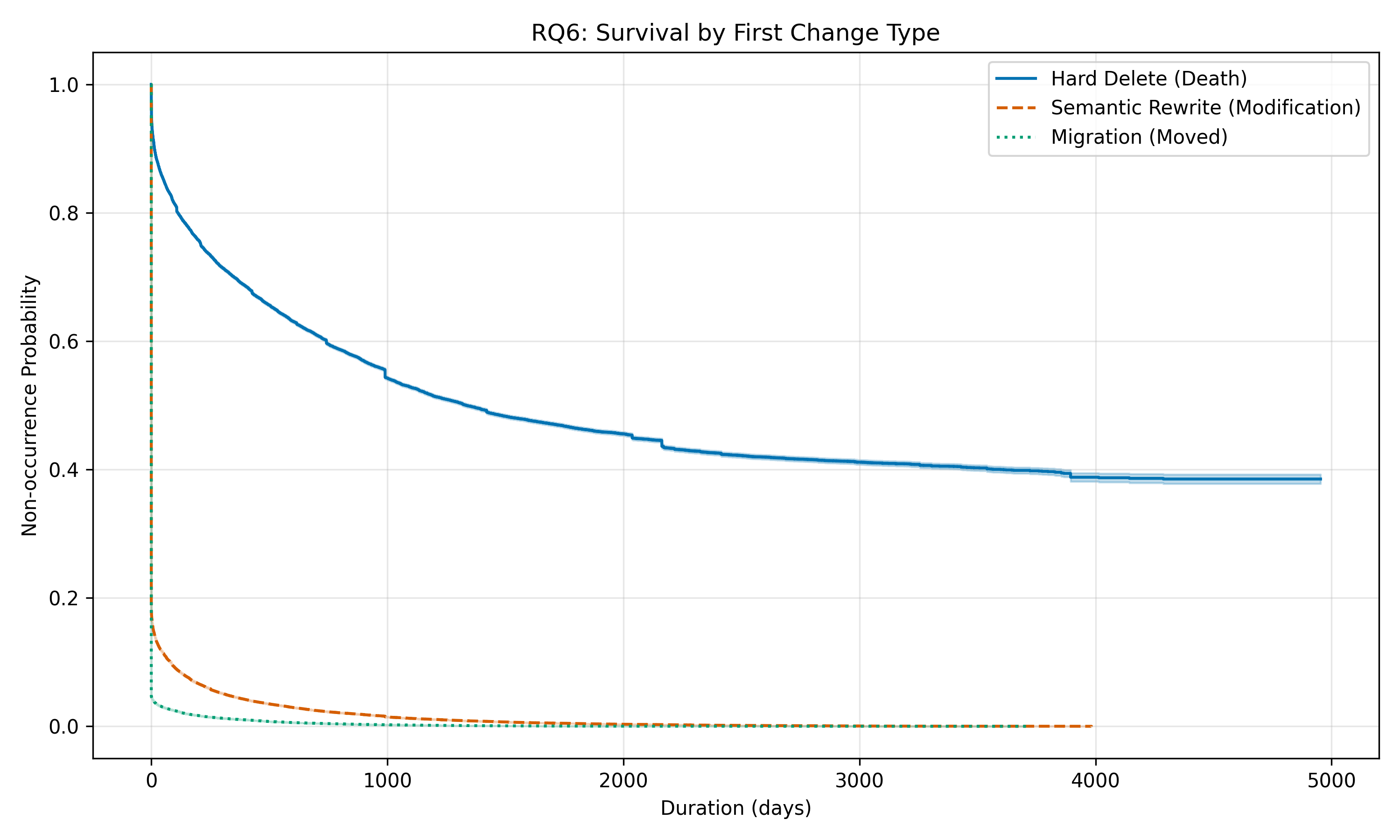}
\caption{Kaplan--Meier curves of time to first change, stratified by
change type: hard deletion (true death), semantic rewrite
(modification), and migration (relocation). The y-axis is the
probability of not having yet undergone that change type. Migrations and
modifications occur almost entirely within the first weeks, whereas hard
deletions have a much heavier right tail. In the survival models only
hard deletions count as events; migrations and modifications are
right-censored.}
\end{figure}

\hypertarget{sensitivity-analysis}{%
\subsubsection{6.7 Sensitivity Analysis}\label{sensitivity-analysis}}

\hypertarget{cox-models-on-data-subsets}{%
\paragraph{6.7.1 Cox Models on Data
Subsets}\label{cox-models-on-data-subsets}}

To evaluate the robustness of findings, we fit four additional Cox
models on subsets of the data:

\begin{enumerate}
\def\labelenumi{\arabic{enumi}.}
\tightlist
\item
  \textbf{Nested-only} (nesting \textgreater{} 0): 205,836 lines, 63,915
  events.
\item
  \textbf{Single repository} (largest repo, \texttt{repo\_id} =
  490315865): 43,844 lines, 10,512 events.
\item
  \textbf{No ephemeral} (duration \textgreater{} 1 day): 323,911 lines →
  subsampled to 300,000, 93,629 events.
\item
  \textbf{Stratified sample} (cap 10,000 lines per repository): 247,572
  lines, 84,974 events. To address the sampling non-stratification
  concern (§8.2), we re-drew the sample with a per-repository cap using
  the same \texttt{cityHash64} ordering for reproducibility, preventing
  large repositories (e.g., microsoft/vscode: ∼43K proportional share →
  capped at 10K) from disproportionately influencing aggregate
  estimates; 13 of 120 repositories were capped (C-index = 0.591).
\end{enumerate}

Table 3 compares hazard ratios across models. We classify covariates as
\textbf{stable} (consistent direction and significance across all
models) or \textbf{unstable} (direction reversal or significance loss).

\begin{longtable}[]{@{}
  >{\raggedright\arraybackslash}p{(\columnwidth - 12\tabcolsep) * \real{0.2800}}
  >{\raggedright\arraybackslash}p{(\columnwidth - 12\tabcolsep) * \real{0.0800}}
  >{\raggedright\arraybackslash}p{(\columnwidth - 12\tabcolsep) * \real{0.1067}}
  >{\raggedright\arraybackslash}p{(\columnwidth - 12\tabcolsep) * \real{0.1067}}
  >{\raggedright\arraybackslash}p{(\columnwidth - 12\tabcolsep) * \real{0.1200}}
  >{\raggedright\arraybackslash}p{(\columnwidth - 12\tabcolsep) * \real{0.1600}}
  >{\raggedright\arraybackslash}p{(\columnwidth - 12\tabcolsep) * \real{0.1467}}@{}}
\toprule\noalign{}
\begin{minipage}[b]{\linewidth}\raggedright
Covariate
\end{minipage} & \begin{minipage}[b]{\linewidth}\raggedright
Main
\end{minipage} & \begin{minipage}[b]{\linewidth}\raggedright
Nested
\end{minipage} & \begin{minipage}[b]{\linewidth}\raggedright
Single
\end{minipage} & \begin{minipage}[b]{\linewidth}\raggedright
NoEphem
\end{minipage} & \begin{minipage}[b]{\linewidth}\raggedright
Stratified
\end{minipage} & \begin{minipage}[b]{\linewidth}\raggedright
Stability
\end{minipage} \\
\midrule\noalign{}
\endhead
\bottomrule\noalign{}
\endlastfoot
\texttt{log\_\discretionary{}{}{}tokens} & 0.87*** & 0.90*** & 0.88*** &
0.88*** & 0.82*** & ✅ Stable \\
\texttt{in\_\discretionary{}{}{}condition} & 1.13*** & 1.17*** & 1.09***
& 1.16*** & 1.18*** & ✅ Stable \\
\texttt{ast\_\discretionary{}{}{}group\_\discretionary{}{}{}expression}
& 1.23*** & 1.24*** & 1.27*** & 1.24*** & 1.26*** & ✅ Stable \\
\texttt{in\_\discretionary{}{}{}function} & 0.81*** & 0.87*** & 0.95* &
0.86*** & 0.80*** & ✅ Stable \\
\texttt{nesting\_\discretionary{}{}{}level} & 0.96*** & 0.98*** &
0.98*** & 0.97*** & 0.97*** & ✅ Stable \\
\texttt{is\_\discretionary{}{}{}night} & 1.16*** & 1.12*** & 1.06* &
1.12*** & 1.10*** & ✅ Stable \\
\texttt{log\_\discretionary{}{}{}todo\_\discretionary{}{}{}distance} &
1.02*** & 0.99*** & 0.97*** & 1.01** & 1.02*** & ❌ Unstable (reversal
in nested and single) \\
\texttt{lidf} & 1.01*** & 1.00 ns & 1.00 ns & 1.02*** & 0.99*** & ❌
Unstable (ns in nested/single; reversal in stratified) \\
\texttt{ast\_\discretionary{}{}{}group\_\discretionary{}{}{}import\_\discretionary{}{}{}export}
& 0.74*** & 0.19 ns & 0.70*** & 0.80*** & 0.68*** & ❌ Unstable (ns/wide
CI nested) \\
\texttt{is\_\discretionary{}{}{}weekend} & 1.00 ns & 0.98 ns & 1.11** &
1.03** & 1.01 ns & ❌ Unstable (risk-increasing in single and
no-ephemeral) \\
\texttt{ast\_\discretionary{}{}{}group\_\discretionary{}{}{}declaration}
& 0.84*** & 0.95 ns & 0.95 ns & 0.87*** & 0.86*** & ❌ Unstable (ns in
nested and single) \\
\end{longtable}

\emph{Table 3: Sensitivity analysis (selected covariates --- those with
at least one noteworthy stability pattern). *** = p \textless{} 0.001,
** = p \textless{} 0.01, * = p \textless{} 0.05, ns = not significant.
The four omitted covariates (\texttt{ast\_group\_other},
\texttt{ast\_group\_type\_system}, \texttt{in\_try\_catch},
\texttt{in\_loop}) are all stable in direction across the four
sensitivity models (nested, single, no-ephemeral, stratified) and are
discussed individually in §6.5 and §6.10. Full per-covariate tables are
available in the replication package.}

\textbf{Stable covariates}: \textbf{log\_tokens},
\textbf{in\_condition}, \textbf{ast\_group\_expression},
\textbf{in\_function}, \textbf{nesting\_level}, and \textbf{is\_night}
maintain consistent direction and significance across all five models.
The stratified per-repository-capped re-draw confirms stability for all
six covariates (HR within roughly 6\% of main-model values, all p
\textless{} 0.001), indicating that the non-stratified hash sampling
does not materially distort the primary findings. The token-count effect
is the most stable lexical predictor by direction, though the
widest-ranging in magnitude: HR 0.82--0.90 across every subset, always p
\textless{} 0.001. Unlike in the main model, \texttt{is\_night} no
longer loses significance in the single-repository subset (HR = 1.06, p
\textless{} 0.05) --- the residual repository-level confounding flagged
in §6.10 attenuates but does not eliminate this effect within a single
project.

\textbf{Unstable covariates}: \texttt{log\_todo\_distance} is
inconsistent: significant and risk-increasing in the main and
no-ephemeral models but significant and \emph{protective} in both the
nested-only and single-repo models --- a direction reversal, not merely
a loss of significance. \texttt{lidf} is significant and risk-increasing
in the main and no-ephemeral models, not significant in the nested-only
or single-repo models, and significant but protective (reversed) in the
stratified model. \texttt{ast\_group\_import\_export} is unreliable in
the nested-only submodel due to the rarity of nested import statements.
\texttt{ast\_group\_declaration} loses significance in both the
nested-only and single-repo models.

\textbf{Simpson's paradox}: \texttt{is\_weekend} is not significant in
the main model (p = 0.67) but becomes a significant risk factor in both
the single-repo (HR = 1.11, p \textless{} 0.01) and no-ephemeral (HR =
1.03, p \textless{} 0.01) models, reflecting confounding by repository-
and duration-specific commit-timing patterns; it remains not significant
in the stratified model (HR = 1.01, ns). The C-index decreases in the
single-repo model (C = 0.559 vs.~0.586 in main), consistent with the
reduced variance available within a single project.

\hypertarget{aft-model-sensitivity-check}{%
\paragraph{6.7.2 AFT Model Sensitivity
Check}\label{aft-model-sensitivity-check}}

To directly address the PH assumption violations identified in §6.8, we
fitted two Accelerated Failure Time (AFT) models --- Weibull AFT and
Log-Logistic AFT --- on the same 300,000-row sample (ridge penalizer λ =
0.1). AFT models make no proportional hazards assumption: instead of
modeling the hazard, they model the logarithm of survival time as a
linear function of covariates. Table 4 reports the resulting time ratios
alongside the Cox hazard ratios, and Table 5 summarizes model fit.

\begin{quote}
\textbf{Interpretation note:} In Cox PH, HR \textless{} 1 indicates
\emph{reduced} hazard (longer survival). In AFT, TR \textgreater{} 1
indicates \emph{longer} survival. The scales are inverted --- a
directionally consistent result means HR \textless{} 1 and TR
\textgreater{} 1, or HR \textgreater{} 1 and TR \textless{} 1.
\end{quote}

\begin{longtable}[]{@{}
  >{\raggedright\arraybackslash}p{(\columnwidth - 8\tabcolsep) * \real{0.2000}}
  >{\raggedright\arraybackslash}p{(\columnwidth - 8\tabcolsep) * \real{0.2000}}
  >{\raggedright\arraybackslash}p{(\columnwidth - 8\tabcolsep) * \real{0.2000}}
  >{\raggedright\arraybackslash}p{(\columnwidth - 8\tabcolsep) * \real{0.2000}}
  >{\raggedright\arraybackslash}p{(\columnwidth - 8\tabcolsep) * \real{0.2000}}@{}}
\toprule\noalign{}
\begin{minipage}[b]{\linewidth}\raggedright
Covariate
\end{minipage} & \begin{minipage}[b]{\linewidth}\raggedright
Cox HR
\end{minipage} & \begin{minipage}[b]{\linewidth}\raggedright
Weibull TR
\end{minipage} & \begin{minipage}[b]{\linewidth}\raggedright
Log-Logistic TR
\end{minipage} & \begin{minipage}[b]{\linewidth}\raggedright
Consistent?
\end{minipage} \\
\midrule\noalign{}
\endhead
\bottomrule\noalign{}
\endlastfoot
Nesting level & 0.963 {[}0.960, 0.966{]}*** & 1.052 {[}1.047,
1.057{]}*** & 1.055 {[}1.050, 1.060{]}*** & ✓ \\
In loop & 0.959 {[}0.925, 0.994{]}* & 1.093 {[}1.035, 1.154{]}** & 1.086
{[}1.026, 1.150{]}** & ✓ \\
In condition & 1.134 {[}1.115, 1.153{]}*** & 0.909 {[}0.886, 0.933{]}***
& 0.940 {[}0.915, 0.966{]}*** & ✓ \\
In try/catch & 1.081 {[}1.036, 1.128{]}*** & 0.954 {[}0.893, 1.019{]}ns
& 0.974 {[}0.910, 1.043{]}ns & ✓ \\
In function & 0.810 {[}0.799, 0.820{]}*** & 1.302 {[}1.279, 1.326{]}***
& 1.322 {[}1.297, 1.347{]}*** & ✓ \\
LIDF (uniqueness) & 1.013 {[}1.008, 1.018{]}*** & 0.996 {[}0.989,
1.003{]}ns & 0.999 {[}0.991, 1.007{]}ns & ✓ \\
Log tokens & 0.872 {[}0.863, 0.881{]}*** & 1.180 {[}1.162, 1.198{]}*** &
1.174 {[}1.155, 1.193{]}*** & ✓ \\
Log TODO distance & 1.015 {[}1.009, 1.021{]}*** & 0.978 {[}0.969,
0.987{]}*** & 0.984 {[}0.974, 0.993{]}*** & ✓ \\
Weekend commit & 1.004 {[}0.986, 1.023{]}ns & 0.993 {[}0.964, 1.022{]}ns
& 0.967 {[}0.938, 0.997{]}* & ✓ \\
Night commit & 1.163 {[}1.145, 1.181{]}*** & 0.822 {[}0.801, 0.843{]}***
& 0.801 {[}0.779, 0.823{]}*** & ✓ \\
AST: declaration & 0.843 {[}0.810, 0.878{]}*** & 1.275 {[}1.203,
1.352{]}*** & 1.265 {[}1.191, 1.345{]}*** & ✓ \\
AST: expression & 1.231 {[}1.193, 1.270{]}*** & 0.831 {[}0.794,
0.869{]}*** & 0.849 {[}0.811, 0.890{]}*** & ✓ \\
AST: import/export & 0.739 {[}0.718, 0.760{]}*** & 1.387 {[}1.337,
1.438{]}*** & 1.412 {[}1.360, 1.466{]}*** & ✓ \\
AST: other & 1.157 {[}1.134, 1.181{]}*** & 0.836 {[}0.815, 0.857{]}*** &
0.824 {[}0.803, 0.845{]}*** & ✓ \\
AST: type system & 0.820 {[}0.740, 0.908{]}*** & 1.241 {[}1.062,
1.449{]}** & 1.226 {[}1.041, 1.443{]}* & ✓ \\
\end{longtable}

\emph{Table 4: Cox HR vs AFT Time Ratios (n = 300,000; ridge penalizer λ
= 0.1; AST reference = \texttt{control\_flow}). *** p \textless{} 0.001,
** p \textless{} 0.01, * p \textless{} 0.05, ns = not significant.
Consistent: HR and TR agree on survival direction (HR \textless{} 1 ↔ TR
\textgreater{} 1). \texttt{in\_try\_catch} is directionally consistent
but loses significance in both AFT models; \texttt{weekend\ commit} is
not significant in Cox or the Weibull AFT but reaches significance in
the Log-Logistic AFT.}

\begin{longtable}[]{@{}lll@{}}
\toprule\noalign{}
Model & AIC & C-index \\
\midrule\noalign{}
\endhead
\bottomrule\noalign{}
\endlastfoot
Weibull AFT & 1,530,022 & 0.584 \\
Log-Logistic AFT & 1,528,471 & 0.584 \\
\end{longtable}

\emph{Table 5: AFT model fit. Log-Logistic AFT achieves lower AIC.
C-index is comparable to the marginal Cox (0.586). BIC values are
omitted: the lifelines library reports BIC within single digits of AIC
for both models (a known issue with BIC computation for parametric AFT),
making them unreliable for comparison.}

\begin{figure}
\centering
\includegraphics{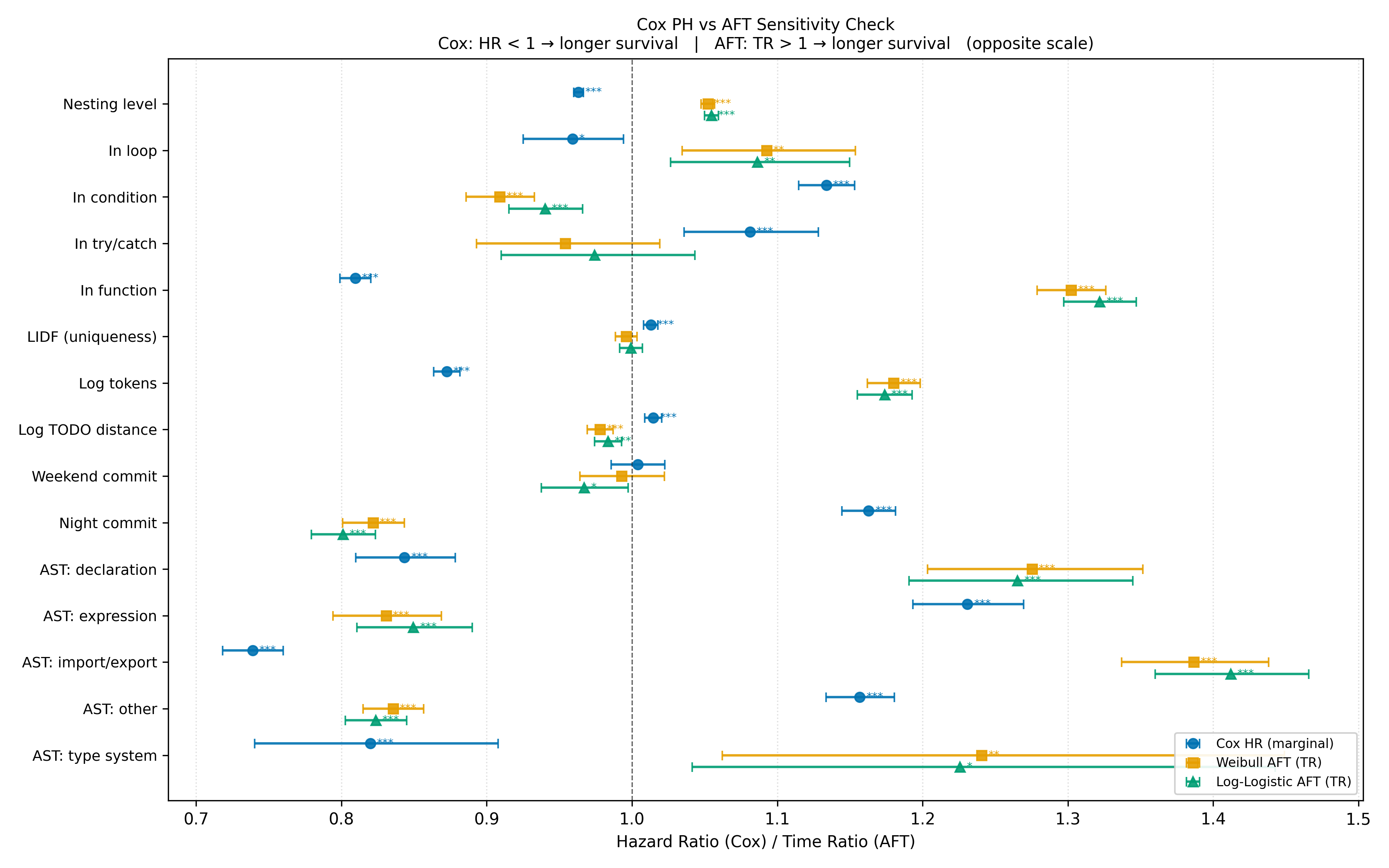}
\caption{Forest plot comparing Cox hazard ratios (circles) with Weibull
AFT (squares) and Log-Logistic AFT (triangles) time ratios, with 95\%
confidence intervals, across all 15 covariates (vertical dashed line at
1.0). Note the inverted scale between the two model families: a Cox HR
\textless{} 1 and an AFT TR \textgreater{} 1 both indicate longer
survival.}
\end{figure}

\textbf{Result:} All 15 of 15 covariates are directionally consistent
between Cox PH and both AFT models. Notably,
\texttt{log\_todo\_distance} (Cox HR = 1.015, risk-increasing) is
confirmed as a risk factor in both AFT models (Weibull TR = 0.978,
Log-Logistic TR = 0.984, both p \textless{} 0.001).
\texttt{in\_try\_catch} is directionally consistent but loses
statistical significance under both AFT models, suggesting a weaker
effect that may be sensitive to distributional assumptions;
\texttt{lidf}, conversely, is significant in the Cox model (HR = 1.013)
but not significant in either AFT model, the reverse pattern. The Cox PH
conclusions are robust to distributional assumptions despite pervasive
PH violations.

\hypertarget{proportional-hazards-diagnostics}{%
\subsubsection{6.8 Proportional Hazards
Diagnostics}\label{proportional-hazards-diagnostics}}

Ten of 15 covariates violate the proportional hazards assumption in the
main model (Schoenfeld residual test, p \textless{} 0.05), confirming
that most, but not all, reported hazard ratios are time-averaged effects
--- a limitation resolved mechanistically via landmark analysis in §6.9
for the covariates that are genuinely time-varying.

\begin{longtable}[]{@{}ll@{}}
\toprule\noalign{}
Model & Covariates Violating PH (p \textless{} 0.05) \\
\midrule\noalign{}
\endhead
\bottomrule\noalign{}
\endlastfoot
Main & 10 / 15 (67\%) \\
Nested-only & 9 / 15 \\
Single repo & 4 / 15 \\
No ephemeral & 10 / 15 \\
\end{longtable}

In the main model, the most severe violations occur for
\texttt{in\_function} (χ² = 191, p ≈ 10⁻⁴³), \texttt{is\_night} (χ² =
159, p ≈ 10⁻³⁶), \texttt{ast\_group\_import\_export} (χ² = 135, p ≈
10⁻³¹), and \texttt{nesting\_level} (χ² = 102, p ≈ 10⁻²⁴). The five
covariates that do not violate PH are \texttt{ast\_group\_expression} (p
= 0.13), \texttt{ast\_group\_type\_system} (p = 0.71),
\texttt{in\_try\_catch} (p = 0.49), \texttt{lidf} (p = 0.21), and
\texttt{log\_tokens} (p = 0.34). The clean Schoenfeld test for token
count is consistent with the essentially flat time-stratified effect
reported in §6.9, and marks a change from an earlier character-entropy
formulation of this feature, whose Schoenfeld statistic was χ² ≈ 474 ---
a formulation-dependent artifact rather than a property of lexical
richness itself. This indicates that most, but not all, hazard ratios
are \emph{not constant over time} --- several represent time-averaged
effects that may overstate or understate the true impact at specific
periods in a line's lifecycle, while a few (notably token count) have
genuinely stable effects.

We interpret Cox results as average effects over the observation period.
As a direct sensitivity check, we fit Weibull AFT and Log-Logistic AFT
models on the same sample (§6.7.2); their conclusions are directionally
consistent with Cox for all 15 of 15 covariates, confirming that PH
violation does not materially distort the reported findings. The
mechanism behind the most striking of these violations,
\texttt{in\_condition}, is resolved in §6.9 via time-stratified
analysis.

\hypertarget{time-varying-effects-landmark-time-stratified-cox}{%
\subsubsection{6.9 Time-Varying Effects: Landmark Time-Stratified
Cox}\label{time-varying-effects-landmark-time-stratified-cox}}

To provide a mechanistic explanation for the PH violations identified in
§6.8 (Proportional Hazards Diagnostics), we fitted separate Cox models
on three landmark time bands: \textbf{0--90 days} (N = 349,510, 58,172
events), \textbf{90--365 days} (N = 228,240, 29,904 events), and
\textbf{365+ days} (N = 140,357, 30,408 events). Unlike the main Cox
model (§6.5, n = 300,000 subsample), these landmark models use the full
349,510-line sample (350,000 minus comment\_like lines); HR estimates
are therefore not directly numerically comparable to Table 2. Each band
uses all lines still at risk at the band's start, with the time origin
shifted to the band boundary and the event indicator re-censored at the
band end. Table 6 reports the per-band hazard ratios for seven
covariates spanning both the time-varying and the (for comparison)
time-stable cases.

\begin{longtable}[]{@{}
  >{\raggedright\arraybackslash}p{(\columnwidth - 12\tabcolsep) * \real{0.2800}}
  >{\raggedright\arraybackslash}p{(\columnwidth - 12\tabcolsep) * \real{0.1200}}
  >{\raggedright\arraybackslash}p{(\columnwidth - 12\tabcolsep) * \real{0.1200}}
  >{\raggedright\arraybackslash}p{(\columnwidth - 12\tabcolsep) * \real{0.1200}}
  >{\raggedright\arraybackslash}p{(\columnwidth - 12\tabcolsep) * \real{0.1200}}
  >{\raggedright\arraybackslash}p{(\columnwidth - 12\tabcolsep) * \real{0.1200}}
  >{\raggedright\arraybackslash}p{(\columnwidth - 12\tabcolsep) * \real{0.1200}}@{}}
\toprule\noalign{}
\begin{minipage}[b]{\linewidth}\raggedright
Covariate
\end{minipage} & \begin{minipage}[b]{\linewidth}\raggedright
HR 0--90d
\end{minipage} & \begin{minipage}[b]{\linewidth}\raggedright
95\% CI
\end{minipage} & \begin{minipage}[b]{\linewidth}\raggedright
HR 90--365d
\end{minipage} & \begin{minipage}[b]{\linewidth}\raggedright
95\% CI
\end{minipage} & \begin{minipage}[b]{\linewidth}\raggedright
HR 365+d
\end{minipage} & \begin{minipage}[b]{\linewidth}\raggedright
95\% CI
\end{minipage} \\
\midrule\noalign{}
\endhead
\bottomrule\noalign{}
\endlastfoot
\texttt{in\_\discretionary{}{}{}condition} & 0.981 ns & {[}0.962,
1.002{]} & 1.229*** & {[}1.198, 1.260{]} & 1.157*** & {[}1.125,
1.190{]} \\
\texttt{log\_\discretionary{}{}{}tokens} & 0.890*** & {[}0.879, 0.901{]}
& 0.905*** & {[}0.891, 0.919{]} & 0.901*** & {[}0.885, 0.917{]} \\
\texttt{ast\_\discretionary{}{}{}group\_\discretionary{}{}{}expression}
& 1.145*** & {[}1.105, 1.186{]} & 1.182*** & {[}1.131, 1.236{]} &
1.177*** & {[}1.118, 1.238{]} \\
\texttt{in\_\discretionary{}{}{}function} & 0.801*** & {[}0.790,
0.813{]} & 0.914*** & {[}0.897, 0.932{]} & 0.842*** & {[}0.824,
0.861{]} \\
\texttt{is\_\discretionary{}{}{}night} & 1.263*** & {[}1.239, 1.287{]} &
1.023 ns & {[}0.996, 1.051{]} & 1.078*** & {[}1.048, 1.108{]} \\
\texttt{nesting\_\discretionary{}{}{}level} & 0.958*** & {[}0.955,
0.962{]} & 0.986*** & {[}0.981, 0.990{]} & 0.977*** & {[}0.972,
0.983{]} \\
\texttt{lidf} & 1.012*** & {[}1.006, 1.018{]} & 1.006 ns & {[}0.999,
1.013{]} & 0.997 ns & {[}0.989, 1.005{]} \\
\end{longtable}

\emph{Table 6: Time-stratified Cox --- HR by landmark band (ridge λ =
0.1). *** p \textless{} 0.001, ns = not significant.}

\begin{figure}
\centering
\includegraphics{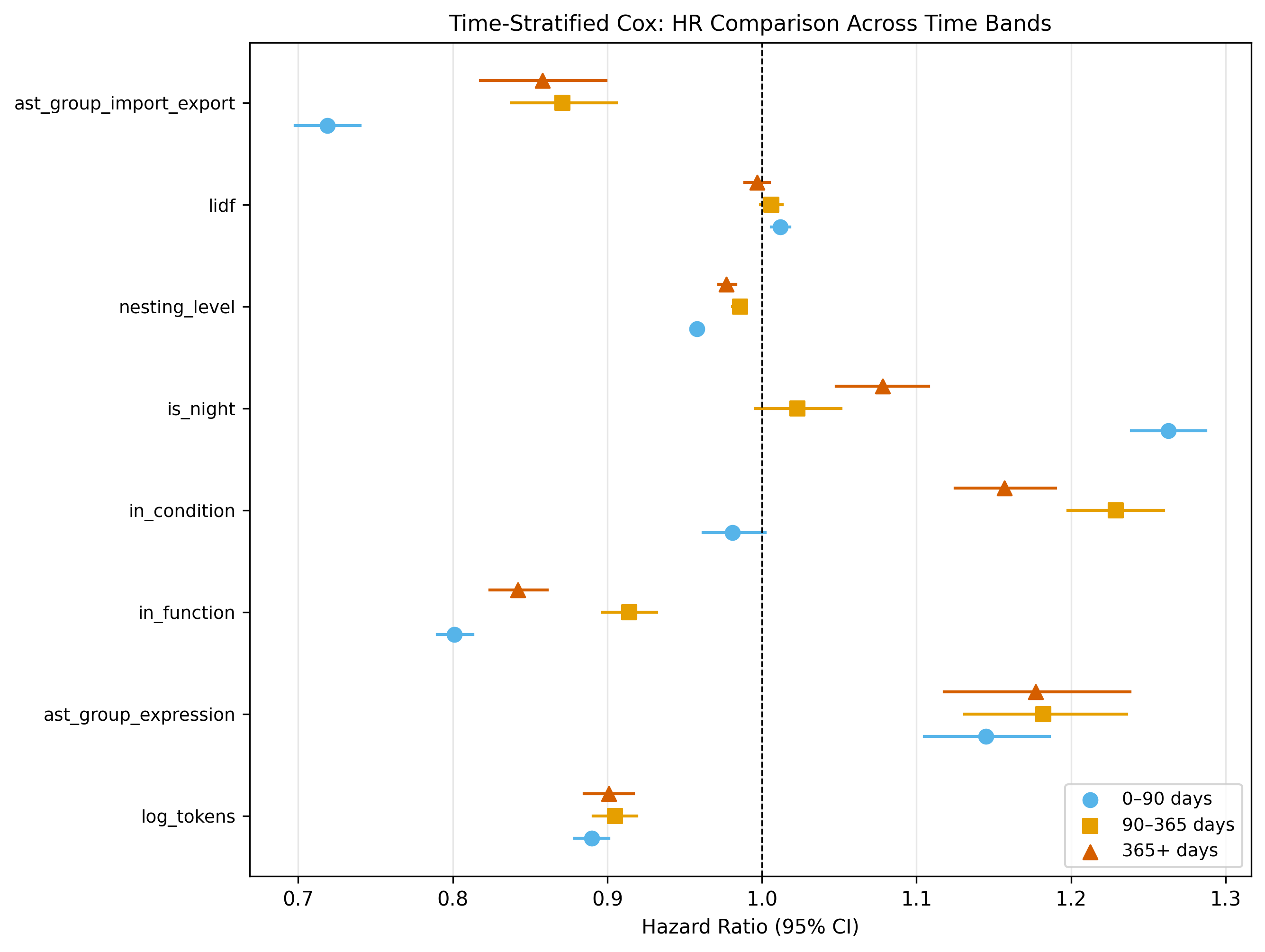}
\caption{Hazard ratios (with 95\% confidence intervals) for seven key
covariates across the three landmark time bands --- 0--90 days
(circles), 90--365 days (squares), and 365+ days (triangles); the
vertical dashed line marks HR = 1. \texttt{in\_condition} has no
detectable effect at birth and becomes risk-increasing after 90 days,
while \texttt{log\_tokens} stays essentially flat across all three
bands.}
\end{figure}

\textbf{Key findings:}

\begin{enumerate}
\def\labelenumi{\arabic{enumi}.}
\item
  \textbf{\texttt{in\_condition} --- from no effect to a risk factor
  (the dominant time-varying effect).} In the 0--90-day band this
  covariate shows no detectable effect (HR = 0.981, p = 0.07, not
  significant), but it becomes a clear risk factor in both the
  90--365-day (HR = 1.229) and 365+-day (HR = 1.157) bands. A young
  conditional branch has not yet become load-bearing, so it neither
  helps nor hurts survival; one that survives past 90 days has entered a
  feature-critical path where it accumulates complexity and attracts
  refactoring. This shift from null to risk-increasing is the clearest
  time-dynamic in the model and the mechanism behind its PH violation
  (§6.8).
\item
  \textbf{\texttt{log\_tokens} --- flat, not time-varying.} The HR is
  0.890 in the early period, 0.905 (90--365d), and 0.901 for code that
  has survived 365+ days --- a difference of at most 1.7 percentage
  points across the entire observation window, well within what sampling
  noise would produce, and consistent with the covariate's clean
  Schoenfeld test (§6.8, p = 0.34, no PH violation). Token-rich lines
  are modestly more durable, by essentially the same amount, at every
  age. We flag this explicitly because an earlier formulation of this
  analysis, using an uncorrected censoring treatment for edited lines,
  reported an artifactual age trend for this same covariate (HR drifting
  from 0.87 to 0.77 with a corresponding spurious Schoenfeld violation);
  correcting that treatment removes the trend entirely, underscoring
  that not every PH violation reflects a genuine mechanism (§6.8).
\item
  \textbf{\texttt{ast\_group\_expression} --- near-time-invariant risk.}
  HRs across the three bands are 1.145, 1.182, and 1.177 --- a small,
  non-monotonic fluctuation around a roughly 15--18\% hazard increase,
  all p \textless{} 0.001. The corresponding Schoenfeld violation is
  weak (χ² = 2.3, p = 0.13 in §6.8, i.e.~not significant on its own) ---
  consistent with expression being best read as an approximately
  time-stable risk factor, included here for contrast with the two
  genuinely time-varying covariates above.
\item
  \textbf{\texttt{is\_night} --- early-period phenomenon.} The
  night-commit risk is concentrated in the 0--90-day band (HR = 1.263)
  and becomes non-significant in the 90--365-day band (HR = 1.023, p =
  0.095). For lines that survive past one year, a modest effect
  resurfaces (HR = 1.078). This pattern indicates that hastily committed
  code is most likely to be revised quickly, but survivors are largely
  similar to daytime code.
\end{enumerate}

These time-resolved estimates sharpen the §7 practical implications:
risk scoring should be \textbf{time-conditional} for the covariates that
are genuinely time-varying, and the dominant such shift is contextual
rather than lexical --- \texttt{in\_condition} moves from no detectable
effect at birth (HR = 0.981, ns) to a risk factor after 90 days (HR =
1.229), whereas token count stays protective and essentially flat
throughout (HR 0.89--0.91). Distinguishing the two cases --- via the
Schoenfeld test in §6.8, not just the landmark table --- is itself part
of the analysis: not every covariate needs age-conditioning, and
assuming otherwise from a PH violation alone can manufacture a time
trend that is not there (point 2 above).

\hypertarget{shared-frailty-model-repository-random-effects}{%
\subsubsection{6.10 Shared Frailty Model: Repository Random
Effects}\label{shared-frailty-model-repository-random-effects}}

\textbf{Motivation.} Two findings from the preceding analysis motivate a
repository-level frailty model: (1) the marginal C-index of
\textbf{0.587} (R \texttt{coxph}, n = 346,808 --- the base for the
frailty comparison; consistent with the Python lifelines estimate of
0.586 reported in §6.5) is low, suggesting substantial unexplained
heterogeneity; and (2) several covariates shift substantially when
conditioning on a single project (§6.7): \texttt{is\_night} attenuates
from HR = 1.163 in the Python marginal model (§6.5) to HR = 1.058 in the
single-repo model, and \texttt{is\_weekend} and \texttt{in\_try\_catch}
change direction or lose significance entirely. Both phenomena are
consistent with strong between-repository confounding: each project
carries its own baseline hazard that, if unmodeled, distorts aggregate
covariate estimates.

\textbf{Model specification.} We fit a shared gamma frailty Cox model
with all 15 fixed-effect covariates and a latent multiplicative cluster
effect per repository:

\[h_i(t \mid Z_i) = Z_i \cdot h_0(t) \cdot \exp(\mathbf{x}_i^\top \boldsymbol{\beta})\]

where \(Z_i \sim \mathrm{Gamma}(1/\theta, 1/\theta)\) is the frailty
term for repository \(i\), with \(\mathbb{E}[Z_i]=1\) and
\(\mathrm{Var}[Z_i]=\theta\). Estimation uses the EM algorithm with
Breslow baseline hazard (R \texttt{survival} package, \texttt{coxph}
with \texttt{frailty(repo\_id,\ distribution="gamma")}). Repositories
with fewer than 30 events are excluded from the frailty clustering, and
comment/shebang lines (\texttt{comment\_like}) are dropped (§6.4). The
final sample comprises \textbf{n = 346,808 observations},
\textbf{118,138 events}.

\textbf{Frailty variance.} The estimated frailty variance
\(\hat{\theta} = 1.451\) (frailty term LRT: \(\chi^2 = 41{,}571\),
\(p \approx 0\)) confirms that repositories are highly heterogeneous in
their baseline hazard. A \(\hat{\theta}\) substantially greater than
zero indicates that between-repository variation is a dominant driver of
observed survival differences, validating the concerns raised under
Simpson's paradox (§6.7) and Internal Validity (§8.2).

\textbf{Discriminative power.} The concordance index improves from 0.587
(marginal Cox) to \textbf{0.676} (frailty model), a gain of +0.089. This
C-index is estimated with known frailty values; for truly out-of-sample
repositories, the effective concordance would be lower.

\textbf{Covariate comparison.} Table 7 presents hazard ratios from both
R models.

Both R models use the same \texttt{ast\_group\ =\ control\_flow}
reference as Table 2 (§6.4), so the AST hazard ratios in Table 7 are
directly comparable to the main Cox model. All within-table (marginal vs
frailty) comparisons are likewise valid.

\textbf{Coefficient scale difference.} The primary source of HR
discrepancy between Table 2 (Python lifelines) and Table 7 (R) is the L2
ridge penalty applied only in the Python model (λ = 0.1), which shrinks
coefficients toward zero (e.g., \texttt{log\_tokens}: Python HR = 0.872
vs R marginal HR = 0.839). Both R models are unpenalized, so the Δ HR
column in Table 7 reflects genuine frailty conditioning rather than a
regularization artifact.

\begin{longtable}[]{@{}
  >{\raggedright\arraybackslash}p{(\columnwidth - 8\tabcolsep) * \real{0.2800}}
  >{\raggedright\arraybackslash}p{(\columnwidth - 8\tabcolsep) * \real{0.2800}}
  >{\raggedright\arraybackslash}p{(\columnwidth - 8\tabcolsep) * \real{0.2800}}
  >{\raggedright\arraybackslash}p{(\columnwidth - 8\tabcolsep) * \real{0.0800}}
  >{\raggedright\arraybackslash}p{(\columnwidth - 8\tabcolsep) * \real{0.0800}}@{}}
\toprule\noalign{}
\begin{minipage}[b]{\linewidth}\raggedright
Covariate
\end{minipage} & \begin{minipage}[b]{\linewidth}\raggedright
Marginal HR {[}95\% CI{]}
\end{minipage} & \begin{minipage}[b]{\linewidth}\raggedright
Frailty HR {[}95\% CI{]}
\end{minipage} & \begin{minipage}[b]{\linewidth}\raggedright
Δ HR
\end{minipage} & \begin{minipage}[b]{\linewidth}\raggedright
Note
\end{minipage} \\
\midrule\noalign{}
\endhead
\bottomrule\noalign{}
\endlastfoot
\textbf{\texttt{in\_condition}} & 1.224 {[}1.200, 1.248{]}*** &
\textbf{1.157 {[}1.134, 1.182{]}}* & −5.5\% & attenuated \\
\textbf{\texttt{is\_night}} & 1.195 {[}1.175, 1.215{]}*** &
\textbf{1.134 {[}1.114, 1.154{]}}* & −5.1\% & Simpson's ↓ \\
\textbf{\texttt{in\_try\_catch}} & 1.154 {[}1.102, 1.208{]}*** &
\textbf{1.051 {[}1.004, 1.100{]} }* & −8.9\% & attenuated, still sig. \\
\texttt{ast\_\discretionary{}{}{}group: expression} & 1.409 {[}1.345,
1.475{]}*** & 1.348 {[}1.287, 1.413{]}*** & −4.3\% & attenuated \\
\texttt{ast\_\discretionary{}{}{}group: other} & 1.243 {[}1.196,
1.291{]}*** & 1.192 {[}1.147, 1.239{]}*** & −4.1\% & attenuated \\
\texttt{nesting\_\discretionary{}{}{}level} & 0.956 {[}0.952,
0.960{]}*** & 0.958 {[}0.953, 0.963{]}*** & +0.2\% & stable \\
\textbf{\texttt{in\_loop}} & 0.957 {[}0.920, 0.996{]}* & \textbf{1.020
{[}0.980, 1.062{]} ns} & LOST & repo proxy \\
\texttt{in\_\discretionary{}{}{}function} & 0.762 {[}0.749, 0.775{]}***
& 0.810 {[}0.796, 0.825{]}*** & +6.3\% & weaker \\
\textbf{\texttt{log\_tokens}} & 0.839 {[}0.830, 0.849{]}*** &
\textbf{0.898 {[}0.888, 0.909{]}}* & +7.0\% & attenuates \\
\textbf{\texttt{log\_todo\_distance}} & 1.016 {[}1.010, 1.022{]}*** &
\textbf{1.024 {[}1.018, 1.031{]}}* & MAINTAINED & remains sig. \\
\texttt{lidf} & 1.024 {[}1.019, 1.030{]}*** & 1.029 {[}1.024,
1.035{]}*** & +0.5\% & weak ↑ \\
\texttt{is\_\discretionary{}{}{}weekend} & 1.021 {[}1.001, 1.041{]}* &
1.014 {[}0.994, 1.035{]} ns & LOST & marginal-only \\
\texttt{ast\_\discretionary{}{}{}group: declaration} & 0.848 {[}0.802,
0.897{]}*** & 0.860 {[}0.812, 0.910{]}*** & +1.4\% & stable \\
\texttt{ast\_\discretionary{}{}{}group: import\_\discretionary{}{}{}export}
& 0.681 {[}0.649, 0.715{]}*** & 0.647 {[}0.617, 0.679{]}*** & −5.0\% &
stronger \\
\texttt{ast\_\discretionary{}{}{}group: type\_\discretionary{}{}{}system}
& 0.773 {[}0.686, 0.871{]}*** & 0.853 {[}0.757, 0.961{]}** & +10.4\% &
weaker \\
\end{longtable}

\emph{Table 7: Marginal Cox vs.~Shared Frailty Cox hazard ratios (R,
unpenalized). Reference: \texttt{ast\_group} = \texttt{control\_flow}
(same as Table 2). *** p \textless{} 0.001, ** p \textless{} 0.01, * p
\textless{} 0.05, ns = not significant. LOST = significance lost after
conditioning; MAINTAINED = remains significant in both models.}

\begin{figure}
\centering
\includegraphics{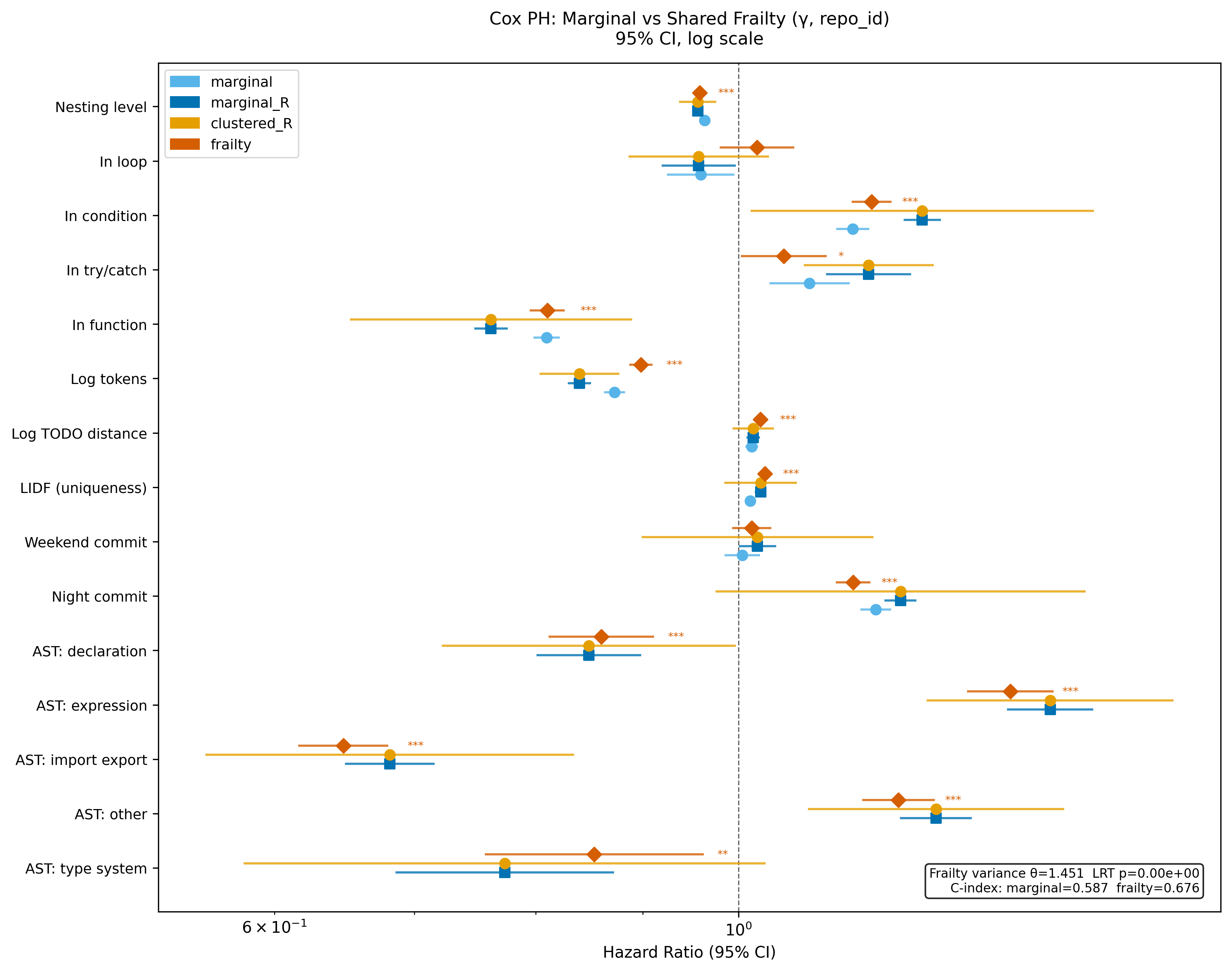}
\caption{Forest plot of hazard ratios (95\% confidence intervals,
log-scaled x-axis) for all 15 covariates under four model variants: the
Python marginal Cox (\texttt{marginal}), the R marginal Cox
(\texttt{marginal\_R}), the R marginal Cox with within-commit clustered
sandwich standard errors (\texttt{clustered\_R}), and the shared gamma
frailty Cox (\texttt{frailty}). The inset reports the frailty variance
(θ = 1.451) and the concordance gain from 0.587 (marginal) to 0.676
(frailty). Risk-side AST groups and \texttt{is\_night} attenuate toward
the null under frailty conditioning, while \texttt{in\_loop} and
\texttt{is\_weekend} lose significance entirely.}
\end{figure}

\textbf{Key findings from frailty conditioning:}

\begin{enumerate}
\def\labelenumi{\arabic{enumi}.}
\item
  \textbf{\texttt{is\_night} --- partial resolution of Simpson's
  paradox.} The nocturnal commit effect is attenuated from HR = 1.195 to
  \textbf{HR = 1.134} (−5.1\%) after conditioning on repository-level
  baseline hazard. The effect remains significant, confirming a genuine
  within-project association; the aggregate marginal estimate was
  partially inflated by between-repository confounding.
\item
  \textbf{\texttt{in\_loop} and \texttt{is\_weekend} --- confirmed
  repository proxies.} Both lose significance entirely under frailty
  conditioning (\texttt{in\_loop}: HR = 1.020, ns; \texttt{is\_weekend}:
  HR = 1.014, ns). Their marginal effects reflected between-repository
  variation in coding style and commit-timing culture, not genuine
  line-level risk. \texttt{in\_try\_catch}, by contrast, attenuates
  substantially (1.154 → 1.051) but remains statistically significant (p
  \textless{} 0.05) --- a partial, not complete, resolution.
\item
  \textbf{\texttt{log\_todo\_distance} --- maintained significance.}
  Significant in both marginal (HR = 1.016) and frailty (HR = 1.024)
  models --- a genuine within-project effect independent of
  repository-level culture.
\item
  \textbf{\texttt{log\_tokens} --- robust but attenuating toward the
  null.} HR 0.839 → 0.898 under frailty conditioning: unlike the
  risk-side groups, the token effect moves \emph{toward} the null rather
  than away from it, but it remains the strongest within-project
  protective lexical predictor (p ≈ 0). \texttt{lidf} (HR = 1.024 →
  1.029, significant in both R models) is a near-null weak risk factor
  after conditioning; the effect size is negligible.
\item
  \textbf{AST risk groups attenuate; protective groups hold.} Relative
  to the control-flow baseline, the risk-side groups move toward the
  null under frailty conditioning --- \texttt{expression} (1.409 →
  1.348) and \texttt{other} (1.243 → 1.192) --- indicating part of their
  aggregate effect reflects repository composition. The protective
  groups are stable (\texttt{declaration} 0.848 → 0.860) or marginally
  stronger (\texttt{import\_export} 0.681 → 0.647), while
  \texttt{type\_system} attenuates toward the null but remains
  significant (0.773*** → 0.853**).
\end{enumerate}

\textbf{Summary.} The shared frailty model confirms that
repository-level heterogeneity (\(\hat{\theta} = 1.451\)) is a dominant
source of variance in code survival. The conditional hazard ratios ---
particularly for \texttt{log\_tokens} (HR = 0.898), \texttt{is\_night}
(HR = 1.134), and \texttt{in\_function} (HR = 0.810) --- represent more
reliable within-project effect estimates than their marginal
counterparts.

\begin{center}\rule{0.5\linewidth}{0.5pt}\end{center}

\hypertarget{practical-implications-and-discussion}{%
\subsection{7. Practical Implications and
Discussion}\label{practical-implications-and-discussion}}

The empirical findings of CLSA carry concrete implications for software
engineering practice, tool development, and future research. We organize
them around three themes.

\hypertarget{ide-risk-scoring}{%
\subsubsection{7.1 IDE Risk Scoring}\label{ide-risk-scoring}}

The strongest and most stable predictors identified in our study ---
\texttt{log\_tokens} (HR = 0.898, frailty model),
\texttt{ast\_group\_expression} (HR = 1.23, main Cox model),
\texttt{in\_function} (HR = 0.810, frailty model), and
\texttt{in\_condition} (HR = 1.157, frailty model) --- are all derivable
statically from a single source file ---
\texttt{ast\_group\_expression}, \texttt{in\_function}, and
\texttt{in\_condition} directly from the AST, and \texttt{log\_tokens}
from the line's tokens --- without any git history. This has an
important practical consequence: these features can be computed at edit
time in an IDE, enabling \emph{live survival scoring} for code being
written.

Concretely, an IDE plugin could use tree-sitter to extract structural
features for each code block and pass them to a survival model trained
on our dataset, displaying an inline risk indicator. For existing lines
with known age from \texttt{git\ blame}, conditional survival S(t
\textbar{} age=T) = S(t+T)/S(T) could adjust the estimate upward for
code that has already survived past the early fragile period (the
\textasciitilde95.7-day median lifespan of deleted lines identified in
§6.3). We leave the implementation and empirical evaluation of such a
tool to future work.

Such a tool operationalizes the ``stabilize or die'' pattern: it gives
developers an actionable risk signal at the moment when it is cheapest
to act --- before the code is committed and reviewed.

\hypertarget{repository-aware-baseline-calibration}{%
\subsubsection{7.2 Repository-Aware Baseline
Calibration}\label{repository-aware-baseline-calibration}}

The dominant predictive signal in our analysis is not any structural
covariate but the repository frailty term (\(\hat{\theta} = 1.451\)),
which accounts for the C-index lift from 0.587 to 0.676. Any practical
risk tool that ignores repository context will be systematically
mis-calibrated: high-turnover projects like \texttt{tamagui/tamagui}
(61\% of mined lines permanently deleted; 71\% among lines with a
resolved outcome) will appear safer than they are when evaluated against
a population-average baseline.

A lightweight calibration approach is feasible without re-training the
model: a cheap repository-level statistic --- the ratio of deleted to
added lines, obtainable from a shallow \texttt{git\ log\ -\/-numstat}
scan --- could serve as a proxy for the project's frailty level and be
used to scale the model's baseline cumulative hazard H₀(t) before
computing survival probabilities, preserving the universally-trained
structural coefficients while adapting the absolute risk level to the
current project. We validated this proxy against the fitted frailty
terms across the 85 repositories retained in the frailty clustering:
Spearman ρ = 0.63 (p ≈ 1×10⁻¹⁰), Pearson r = 0.54. The correlation is
far from perfect and a fuller calibration study --- including comparison
against alternative shallow proxies --- is left to future work, but it
is strong enough to suggest that project-level risk can be approximated
without access to the full historical dataset used to fit the frailty
model.

\hypertarget{code-review-and-cicd-integration}{%
\subsubsection{7.3 Code Review and CI/CD
Integration}\label{code-review-and-cicd-integration}}

In continuous integration workflows, reviewers face a triage problem:
which changed lines deserve the most scrutiny? Our model provides a
direct answer. Lines with high predicted hazard --- expression-type AST
nodes, few tokens (short, lexically simple lines), and established
\texttt{in\_condition} context --- are statistically more likely to be
refactored within the next quarter. An automated pull-request annotator
could flag such lines for focused review, reducing the cognitive load on
human reviewers who currently must apply this judgment heuristically.

This framing is distinct from defect prediction (which requires
historical bug data): survival scoring requires only static features of
the code being submitted, making it applicable to any new codebase
without a pre-existing fault history.

\begin{center}\rule{0.5\linewidth}{0.5pt}\end{center}

\hypertarget{threats-to-validity}{%
\subsection{8. Threats to Validity}\label{threats-to-validity}}

\hypertarget{construct-validity}{%
\subsubsection{8.1 Construct Validity}\label{construct-validity}}

\begin{itemize}
\item
  \textbf{Semantic equivalence}: Our 5-stage matching pipeline uses
  composite textual similarity (Sørensen--Dice + Ratcliff/Obershelp,
  threshold ≥ 0.6) constrained by AST node type agreement (relaxed to ≥
  0.9 in Stage 5) to approximate semantic equivalence. This is a
  necessary heuristic: a line classified as ``migrated'' may represent a
  semantically distinct construct that happens to share textual
  similarity, while a genuinely new line may coincidentally match a
  deleted line's normalized form, producing a false negative in death
  detection. \textbf{Threshold sensitivity.} To quantify the effect of
  the 0.6 choice, we re-ran the full matching pipeline (stages 1--5) on
  three held-out repositories (markedjs/marked,
  piotrwitek/react-redux-typescript-guide, unform/unform; 275 commits
  total) at T = 0.50, 0.60, and 0.70. Results are shown in Table 8.

  \begin{longtable}[]{@{}lrrr@{}}
  \toprule\noalign{}
  Repository & T = 0.50 deaths & T = 0.60 deaths & T = 0.70 deaths \\
  \midrule\noalign{}
  \endhead
  \bottomrule\noalign{}
  \endlastfoot
  markedjs/\discretionary{}{}{}marked & 440 & 480 & 537 \\
  piotrwitek/\discretionary{}{}{}react-redux-typescript-guide & 391 &
  458 & 535 \\
  unform/unform & 1,283 & 1,362 & 1,442 \\
  \end{longtable}

  \emph{Table 8. Death event counts under three similarity thresholds.
  Rows are held-out repositories not used in main model fitting.}

  Lowering the threshold to 0.50 reduces death counts by 6--15\% (avg.
  −9.6\%) relative to T = 0.60; raising it to 0.70 increases death
  counts by 6--17\% (avg. +11.5\%). Manual inspection of the 30
  borderline pairs (0.55 ≤ sim ≤ 0.70) that flip classification between
  thresholds confirms that the T = 0.60 boundary is well-placed: pairs
  promoted to ``modification'' only at T = 0.50 are semantically
  distinct (e.g.,
  \texttt{import\ \{\ defineConfig\ \}\ from\ \textquotesingle{}tsup\textquotesingle{}}
  matched against
  \texttt{import\ type\ \{\ ResultCallback\ \}\ from\ \textquotesingle{}./Instance.ts\textquotesingle{}},
  sim = 0.58), while pairs reclassified as ``death'' only at T = 0.70
  share clear semantic intent (e.g.,
  \texttt{let\ ret\ =\ hooksFunc.apply(hooks,\ args)}
  vs.~\texttt{let\ ret\ =\ pack.renderer!{[}prop{]}.apply(renderer,\ args)},
  sim = 0.67). The ±10\% swing in death event volume, on a
  full-population sample of 350,000 lines, would produce at most a
  sub-percent change in Cox hazard ratios through pure dilution of the
  event pool; directional conclusions are therefore robust to this
  threshold choice.
\item
  \textbf{Death vs.~refactoring}: Despite the multi-stage matching
  pipeline, some genuine line evolutions may be misclassified as deaths
  (or vice versa). Our manual validation of 200 stratified lines (§8.1)
  provides empirical evidence of pipeline reliability (precision =
  1.000, F1 = 0.870).
\item
  \textbf{Line as unit of analysis}: We treat individual lines as
  independent subjects, but lines within the same function or file are
  not statistically independent. Block-level deletions (e.g., removing
  an entire function) cause correlated deaths that violate the
  independence assumption. Frailty models or clustered survival
  approaches could address this.
\item
  \textbf{Censoring mechanism}: We use the repository's last commit date
  as the censoring time. For lines born near the end of the observation
  window, this produces short follow-up times that may systematically
  differ from lines with longer observation periods (informative
  censoring). We partially address this by excluding ephemeral lines
  (\textless1 day) in sensitivity analysis.
\item
  \textbf{Project dormancy and the ``undead'' state}: Because a line is
  recorded as dead only on explicit deletion, a repository whose
  development ceases does not kill its lines --- they are right-censored
  at that repository's last commit rather than counted as events. A
  frozen but still-depended-upon package therefore enters the dataset as
  a population of long-lived survivors. This is the correct survival
  semantics, but it differs from an intuitive ``the code died because
  its project died'' reading, and high survivorship in low-activity
  repositories should be interpreted accordingly. Because censoring is
  anchored to each repository's \emph{own} last commit rather than a
  single global snapshot date, dormancy does not artificially extend
  observed lifespans beyond the point of freeze (cf.~the truncation-bias
  discussion in §8.3). We owe the ``survival to the undead state''
  framing of this behavior to correspondence with a colleague.
\item
  \textbf{Informative censoring from modification and migration}: Lines
  classified as \emph{modifications} (similarity ≥ 0.6) or
  \emph{migrations} are right-censored at the moment of their
  transformation rather than at the repository's last commit. Standard
  survival analysis requires that censoring be non-informative
  (independent of the hazard of the true event). However, if modified or
  migrated lines systematically differ in their structural
  characteristics from lines that undergo true deletion --- which is
  likely, since lexically dense, deeply-nested lines may be more prone
  to in-place modification than outright removal --- then this censoring
  is informative. Table 1 shows that the migration/modification gap
  (Lines Mined − Deaths − Survived) reaches 28.7\% for microsoft/vscode
  and is non-trivial across most large repositories. Quantifying the
  direction and magnitude of this bias would require a formal competing
  risks analysis (Fine--Gray model), which is deferred to future work.
  This event definition also differs deliberately from Spinellis et
  al.~{[}25{]}, who count any non-whitespace modification as a line
  death; we treat in-place modification as censoring rather than an
  event, on the view that a rewritten line has not been \emph{removed}.
  Our definition is the more conservative of the two and avoids
  conflating editing with deletion, at the cost of the
  informative-censoring risk noted above.
\item
  \textbf{Diff-algorithm sensitivity}: Our line identity tracking is
  built on PyDriller's default Myers diff. Spinellis et al.~{[}25{]}
  report that the choice of diff algorithm has a negligible effect on
  fine-grained line-lifetime measurement --- a move-aware Heckel diff
  changes the line set by only ≈ 2.2\% relative to Myers, and a
  Histogram diff differs by \textless{} 0.5\% --- so we do not expect
  our conclusions to be sensitive to this choice. Our 5-stage matching
  pipeline (§5.2) additionally recovers moved and renamed lines that a
  plain Myers diff would misclassify as death + birth.
\item
  \textbf{Structural context extraction}: The four binary structural
  context features (\texttt{in\_loop}, \texttt{in\_condition},
  \texttt{in\_try\_catch}, \texttt{in\_function}) are derived from
  tree-sitter's parent-chain traversal. In languages with complex
  scoping (e.g., nested arrow functions inside conditionals), these
  features may not perfectly capture the programmer's intended
  structural context.
\item
  \textbf{Robustness of Survival Constructs (Manual Validation)}: To
  evaluate construct validity, we manually inspected a stratified random
  sample of 200 lines: 70 \texttt{hard\_delete} deaths, 30
  \texttt{file\_delete} deaths, and 100 censored (surviving) lines. For
  each line, one author inspected the corresponding GitHub commit diff
  and recorded a ground-truth verdict (TP/FP/TN/FN).

  \begin{longtable}[]{@{}lll@{}}
  \toprule\noalign{}
  & Pipeline: DEAD & Pipeline: ALIVE \\
  \midrule\noalign{}
  \endhead
  \bottomrule\noalign{}
  \endlastfoot
  \textbf{Human: dead} & TP = 100 & FN = 30 \\
  \textbf{Human: alive} & FP = 0 & TN = 70 \\
  \end{longtable}

  \begin{longtable}[]{@{}ll@{}}
  \toprule\noalign{}
  Metric & Value \\
  \midrule\noalign{}
  \endhead
  \bottomrule\noalign{}
  \endlastfoot
  Precision & 1.000 \\
  Recall & 0.769 \\
  F1 & 0.870 \\
  Accuracy & 0.850 \\
  Oracle agreement & 0.700 \\
  \end{longtable}

  The pipeline achieves \textbf{precision of 1.000} (0 false positives
  among 100 predicted deaths) and \textbf{recall of 0.769} (F1 = 0.870,
  accuracy = 0.850). We additionally report an oracle-agreement rate of
  0.700, computed as the fraction of the \textbf{100 censored
  (pipeline-alive) observations} for which the pipeline verdict matches
  the human verdict (TN / total\_censored = 70/100 = 0.700). Note: the
  overall accuracy across all 200 lines is (TP + TN)/200 = 170/200 =
  0.850, reported separately in the Accuracy row above. Cohen's κ in its
  canonical form requires two independent human raters and does not
  apply here since a single author performed all annotations.

  All 30 false negatives originate from the censored stratum. Inspection
  of annotator notes reveals the following breakdown: 14 cases where the
  tracked file no longer exists at HEAD, 6 cases where the file exists
  but the specific line content was absent, 5 cases where the string
  content could not be matched, 2 cases of missing parent directories,
  and 2 cases of explicit file renames not propagated into
  \texttt{file\_state}; one case lacked annotator notes. In total,
  \textbf{18 of 30 FN cases} are attributable to the tracked path (file
  or directory) no longer existing at HEAD --- predominantly because
  Git's rename heuristic failed to detect directory-level restructuring,
  leaving lines alive in \texttt{file\_state} under a stale path. These
  findings are consistent with the matching pipeline design (§5.2) and
  motivated the cross-file structural matching (Stage 4) added to
  mitigate this class of errors. The \textbf{zero false-positive rate}
  confirms that the pipeline's conservative design --- preferring to
  leave a line alive when a plausible match exists --- does not produce
  spurious deaths.

  \textbf{Bias direction from recall = 0.769.} The 30.0\% false-negative
  rate among censored observations implies a systematic upward bias in
  the Kaplan--Meier survival estimate: lines that are truly dead but
  recorded as censored inflate the apparent survivorship at every time
  point. The consequence for Cox hazard ratios is an attenuation toward
  HR = 1.0 --- covariates associated with higher deletion risk will
  appear weaker than their true effect, because a fraction of their
  ``events'' are absorbed into the censored pool. The five most robust
  covariates in Table 3 (\texttt{log\_tokens}, \texttt{in\_condition},
  \texttt{ast\_group\_expression}, \texttt{in\_function},
  \texttt{nesting\_level}) are therefore conservative estimates; the
  true hazard ratios for these features are likely further from 1.0 than
  reported.
\end{itemize}

\hypertarget{internal-validity}{%
\subsubsection{8.2 Internal Validity}\label{internal-validity}}

\begin{itemize}
\tightlist
\item
  \textbf{Within-commit correlation.} Lines removed in the same commit
  (e.g., bulk function deletion) share a death event, violating the Cox
  independence assumption. We addressed this by fitting a marginal Cox
  model with clustered sandwich standard errors by \texttt{commit\_id}
  (107,802 unique commits; R \texttt{coxph} with
  \texttt{cluster(commit\_id)}). CI widening is substantial --- 9.27×
  for \texttt{in\_condition}, 8.89× for \texttt{in\_function}, 4.22× for
  \texttt{nesting\_level}, 3.87× for \texttt{log\_tokens}, and 2.82× for
  \texttt{ast\_group\_expression} --- reflecting high within-commit
  correlation (mean ≈ 39 deaths per commit, with the majority of
  deletions concentrated in a small number of very large commits).
  Nonetheless, the stable predictors remain statistically significant
  under sandwich SE (\texttt{log\_tokens} p \textless{} 0.001,
  \texttt{in\_condition} p = 0.041, \texttt{in\_function} p \textless{}
  0.001, \texttt{nesting\_level} p = 0.003,
  \texttt{ast\_group\_expression} p \textless{} 0.001).
  \texttt{is\_night} (p = 0.053) and \texttt{in\_loop} (p = 0.23) lose
  significance under clustering and should be considered unreliable for
  within-commit inference. Directional conclusions for the main
  predictors are robust to this correction.
\item
  \textbf{Commit granularity and merge policy.} Mainline commit
  granularity is not a faithful record of how individual developers
  commit. Under the squash-on-merge workflow that is now common in the
  GitHub-flow JavaScript/TypeScript ecosystem, an entire feature branch
  --- however the author chose to commit on it --- collapses into a
  single mainline commit at merge time. The commits we observe therefore
  reflect a project's \emph{merge policy} far more than any individual's
  commit cadence, and we make no attempt to recover or model developer
  commit style as a covariate.
\item
  \textbf{Proportional Hazards assumption violation}: 10 of 15
  covariates violate the PH assumption in the main model (Schoenfeld
  test). The reported hazard ratios for these represent time-averaged
  effects. The most severe violations occur for \texttt{in\_function}
  (χ² = 191), \texttt{is\_night} (χ² = 159),
  \texttt{ast\_group\_import\_export} (χ² = 135), and
  \texttt{nesting\_level} (χ² = 102). We address this via AFT
  sensitivity analysis (§6.7.2) --- all 15 covariates are directionally
  consistent with Cox PH --- and via landmark time-stratified Cox
  analysis (§6.9), which reveals the mechanistic trajectory of the
  dominant dynamic, \texttt{in\_condition} (from no effect to
  risk-increasing: 0.981 (ns) → 1.229 → 1.157 across 0--90/90--365/365+
  day bands), and confirms that \texttt{log\_tokens} --- despite a large
  PH violation in an earlier formulation of this analysis --- is in fact
  flat across all three bands and does not violate PH once its censoring
  treatment is corrected (§6.9).
\item
  \textbf{Simpson's paradox}: Several covariates lose significance or
  shift direction in sensitivity analysis, indicating repository-level
  confounding. We address this in §6.10 via a shared gamma frailty Cox
  model (\(\hat{\theta} = 1.451\), LRT χ² = 41,571, p ≈ 0).
  \texttt{in\_loop} and \texttt{is\_weekend} lose significance entirely
  under frailty conditioning and are confirmed as proxies for
  repository-level patterns rather than genuine line-level predictors;
  \texttt{in\_try\_catch} attenuates substantially but remains
  significant.
\item
  \textbf{C-index and practical significance}: The limited
  discriminative power (C = 0.586) indicates that structural features
  alone are insufficient to predict code survival. The frailty model
  C-index (0.676) confirms that repository-level information is the
  dominant predictive signal. All reported covariates with HR close to
  1.0 (e.g., \texttt{lidf}, \texttt{in\_try\_catch}) should be
  interpreted with caution.
\item
  \textbf{Temporal confounding}: The \texttt{is\_night} and
  \texttt{is\_weekend} covariates are derived from commit timestamps. In
  cases involving cloud IDEs, remote servers, or misconfigured clocks,
  the timestamp may not reflect the committer's actual timezone; we
  treat this as non-systematic noise. A more systematic concern, raised
  in correspondence, is squash-on-merge: when a feature branch is
  squashed into a single commit at merge time, a line's recorded birth
  commit carries the \emph{merge} timestamp rather than the moment the
  code was authored. The time-of-day label then reflects when the pull
  request landed (often CI- or reviewer-driven) rather than when the
  line was written --- a particularly relevant effect in the GitHub-flow
  JavaScript/TypeScript ecosystem, where squash-merge is common.
  Importantly, this mechanism would \emph{attenuate} a genuine
  authoring-time effect toward the null rather than manufacture one, so
  the residual \texttt{is\_night} effect that survives frailty
  conditioning (HR = 1.126, §6.10) should be read as a noisy,
  repository-confounded signal whose literal time-of-day interpretation
  we do not rely on.
\item
  \textbf{Sampling strategy}: Our deterministic hash-based sampling
  (\texttt{ORDER\ BY\ cityHash64(id)\ LIMIT\ N}) produces a reproducible
  simple random sample but does not stratify by repository. Large
  projects (vscode: 4M lines) are proportionally overrepresented in the
  350K sample, potentially amplifying repository-level confounding
  already noted under Simpson's paradox. We address this directly in
  §6.7 via a stratified sensitivity model (cap 10,000 lines per
  repository, N = 247,572), which confirms that all five primary
  findings are robust to this sampling choice (HR estimates within 4\%
  of main model values).
\end{itemize}

\hypertarget{external-validity}{%
\subsubsection{8.3 External Validity}\label{external-validity}}

\begin{itemize}
\tightlist
\item
  \textbf{Language scope}: Results are constrained to TypeScript, a
  dynamically-typed language with rapid ecosystem evolution.
  Strongly-typed, compiled languages (C++, Rust, Java) with stricter
  type systems and longer release cycles may exhibit fundamentally
  different survival dynamics. The non-significance of
  \texttt{type\_system} AST nodes in the expanded model may be specific
  to TypeScript's gradual typing paradigm.
\item
  \textbf{Repository selection bias}: We exclusively mined active
  open-source repositories from GitHub. Dead or archived repositories,
  private/corporate codebases, and projects hosted on other platforms
  (GitLab, Bitbucket) are not represented. The survival characteristics
  of code in corporate monorepos with structured code review processes
  may differ significantly.
\item
  \textbf{Temporal scope}: Our repositories span 6 to 163 months of
  history. The TypeScript ecosystem evolved substantially over this
  period (e.g., introduction of strict mode, template literal types,
  satisfies operator). Cohort effects---where lines born in different
  eras face different hazard environments---are not explicitly modeled.
  Furthermore, the observation window heterogeneity creates a truncation
  bias: lines in 6-month-old repositories cannot exhibit survival
  durations beyond 6 months, making their censoring rate structurally
  higher than in mature repositories. This systematically inflates the
  apparent survivorship contributed by short-lived repositories. Future
  analyses should apply a minimum repository age filter (e.g., ≥ 24
  months) and verify that key findings hold within that restricted
  cohort.
\item
  \textbf{Project domain}: Our repository selection spans diverse
  domains but is biased toward web development and tooling (the natural
  habitat of TypeScript). Results may not generalize to other software
  domains.
\end{itemize}

\hypertarget{statistical-validity}{%
\subsubsection{8.4 Statistical Validity}\label{statistical-validity}}

\begin{itemize}
\tightlist
\item
  \textbf{Multiple testing}: Benjamini--Hochberg FDR correction is
  applied to all reported log-rank p-values. We distinguish raw from
  adjusted significance throughout. Cox model p-values are reported
  as-is from the partial likelihood ratio test. Two comparisons
  (\texttt{control\_flow\ vs\ type\_system},
  \texttt{declaration\ vs\ type\_system}) lose significance after FDR
  correction.
\item
  \textbf{Large-sample significance}: With 300,000+ observations, even
  trivially small effects achieve statistical significance. We report
  effect sizes (HR) and confidence intervals alongside p-values, and
  explicitly note covariates with HR close to 1.0 (e.g.,
  \texttt{in\_loop} HR = 0.96, \texttt{lidf} HR = 1.01,
  \texttt{is\_weekend} HR = 1.00) as statistically negligible or
  borderline.
\item
  \textbf{L2 regularization}: We apply an L2 penalty (0.1) to the Cox
  model to prevent overfitting. While this shrinks coefficients toward
  zero and may attenuate true effects, it guards against spurious
  associations in high-dimensional settings. The penalty was not tuned
  via cross-validation.
\item
  \textbf{Borderline covariates}: 14 of the 15 covariates are
  statistically significant (p \textless{} 0.05) in the main model; only
  \texttt{is\_weekend} (p = 0.67) is not. Of the significant ones,
  \texttt{in\_loop} (p = 0.023) would not survive Bonferroni correction
  for 15 comparisons (α/15 ≈ 0.003) and should be treated as borderline;
  it also loses significance under both clustered SE and frailty
  conditioning.
\end{itemize}

\hypertarget{reproducibility}{%
\subsubsection{8.5 Reproducibility}\label{reproducibility}}

\begin{itemize}
\tightlist
\item
  All analysis scripts, database schemas (ClickHouse DDL), and raw
  output files are included in the supplementary material. The sampling
  strategy uses deterministic hashes, ensuring exact reproducibility
  given the same database state. The Git mining pipeline (PyDriller +
  tree-sitter) is deterministic given fixed repository snapshots. We
  provide SHA hashes for all repository snapshots used.
\end{itemize}

\begin{center}\rule{0.5\linewidth}{0.5pt}\end{center}

\hypertarget{conclusion}{%
\subsection{9. Conclusion}\label{conclusion}}

In the present work, we establish that survival analysis provides a
statistically coherent and empirically productive framework for modeling
the persistence of individual source code lines. Our primary findings,
ordered by novelty and practical significance, are:

\begin{enumerate}
\def\labelenumi{\arabic{enumi}.}
\item
  \textbf{Code survival is time-varying for some covariates, but not
  all}: Landmark time-stratified Cox analysis (§6.9) shows that
  covariate effects can differ substantially across three time bands
  (\textless{} 90d, 90--365d, 365+d) --- but do not always.
  \texttt{in\_condition} is the most striking time-varying example: no
  detectable effect in new code (HR = 0.981, 0--90d, not significant)
  but a clear risk factor thereafter (HR = 1.229 at 90--365d, HR = 1.157
  at 365+d). Line token count, by contrast, is essentially flat across
  all three bands (HR 0.89--0.91) and shows no proportional-hazards
  violation. Distinguishing genuinely time-varying covariates from
  stable ones --- rather than assuming every PH violation reflects an
  age-dependent mechanism (§6.8) --- is itself part of the empirical
  basis for time-conditional risk scoring.
\item
  \textbf{Repository heterogeneity is the dominant predictive signal}:
  The shared gamma frailty model (§6.10) estimates
  \(\hat{\theta} = 1.451\) (LRT χ² = 41,571, p ≈ 0), confirming that
  between-repository variation dominates cross-project survival
  differences. C-index improves from 0.587 (marginal Cox) to 0.676
  (frailty). Any risk model that ignores project-level context will be
  systematically mis-calibrated. Frailty-conditional hazard ratios
  should be preferred for within-project applications.
\item
  \textbf{Line token count is a robust protective lexical predictor}:
  Lexically richer lines show a 12.8\% hazard reduction (HR = 0.87, main
  model), stable across all sensitivity checks, all three time bands
  (flat, no PH violation), and both AFT models. Crucially, token count
  is \emph{not} reducible to character length --- it supersedes line
  Shannon entropy, which we found to be a line-size proxy (§6.5) --- and
  together with import/export and \texttt{in\_function} lines it is
  among the strongest statically-computable protective factors.
\item
  \textbf{Structural context features are stable and actionable}:
  \texttt{in\_function} is a strong, time-stable protective predictor
  (HR = 0.762--0.810 across marginal and frailty R models).
  \texttt{ast\_group\_expression} is the most stable risk factor (HR =
  1.145--1.182 across all three time bands). Both are computable
  statically from the AST without git history, enabling real-time
  structural risk scoring. By contrast, \texttt{in\_loop} and
  \texttt{is\_weekend} are confirmed as repository-level proxies with no
  genuine line-level predictive value after frailty conditioning;
  \texttt{in\_try\_catch} attenuates substantially under the same
  conditioning but remains statistically significant.
\item
  \textbf{The 5-stage pipeline is a prerequisite, not a preprocessing
  detail}: The bipartite matching pipeline prevented 8.3 million false
  deaths --- 43\% of all pipeline-processed deletion events (8.3M false
  + 11.0M confirmed = 19.3M total observed by the pipeline) --- in the
  full population. Without it, aggregate hazard estimates would be
  severely inflated and covariate effects distorted. The pipeline's
  conservative design (precision = 1.000 in manual validation) ensures
  that reported hazard ratios are biased toward null rather than away
  from it.
\item
  \textbf{PH violations are informative, not uniform}: 10 of 15
  covariates violate the proportional hazards assumption (Schoenfeld
  test); the other 5 --- including, notably, line token count --- do
  not. Rather than treating every violation as a limitation to be
  explained away, the time-stratified analysis (§6.9) shows that the
  violating covariates include both genuine mechanistic dynamics
  (\texttt{in\_condition}) and, in an earlier formulation of the
  analysis, an artifact of an uncorrected censoring treatment (§6.9,
  point 2) that produced a spurious violation for token count. Weibull
  and Log-Logistic AFT models confirm directional consistency for all 15
  covariates, establishing that PH violations affect the magnitude and
  shape of effects over time, but not their direction.
\end{enumerate}

We strictly avoid causality claims. CLSA delivers an infrastructure for
quantifying code evolution patterns empirically --- a foundation for
time-conditional risk scoring tools and a reusable analytical pipeline
for future MSR studies extending beyond TypeScript. Full reproduction
scripts, database schemas, and raw output are available in the
supplementary material.

\begin{center}\rule{0.5\linewidth}{0.5pt}\end{center}

\hypertarget{declarations}{%
\subsection{Declarations}\label{declarations}}

\textbf{Data Availability Statement} The datasets generated and/or
analyzed during the current study, including the ClickHouse database
schema, Python data extraction scripts (PyDriller and tree-sitter
configurations), and full replication pipelines, are openly available on
Zenodo at https://doi.org/10.5281/zenodo.20367614. A curated random
sample used for manual validation is also provided alongside the
analysis scripts to ensure full reproducibility.

\textbf{Acknowledgments} The author thanks the open-source maintainers
whose repositories constitute the CLSA dataset. No external funding was
received for this work.

The author further thanks a correspondent for pointing out closely
related prior work and for several framing suggestions reflected in §8.1
and §8.2.

\textbf{Conflict of Interest} The author declares that they have no
competing interests.

\begin{center}\rule{0.5\linewidth}{0.5pt}\end{center}

\hypertarget{references}{%
\subsection{References}\label{references}}

{[}1{]} J. R. Falleri, F. Morandat, X. Blanc, M. Martinez, and M.
Monperrus, ``Fine-grained and accurate source code differencing,'' in
\emph{Proc. 29th ACM/IEEE Int. Conf. on Automated Software Engineering
(ASE)}, 2014, pp.~313--324.

{[}2{]} D. Silva and M. T. Valente, ``RefDiff: Detecting Refactorings in
Version Histories,'' in \emph{Proc. MSR}, 2017, pp.~269--279.

{[}3{]} E. N. Samoladas, L. Angelis, and I. Stamelos, ``Survival
analysis on the duration of open source projects,'' \emph{Information
and Software Technology}, vol.~52, no. 9, pp.~902--922, 2010.

{[}4{]} D. Spadini, M. Aniche, and A. Bacchelli, ``PyDriller: Python
framework for mining software repositories,'' in \emph{Proc. ESEC/FSE},
2018, pp.~908--911.

{[}5{]} S. Raemaekers, A. van Deursen, and J. Visser, ``Semantic
versioning versus breaking changes: A study of the maven repository,''
in \emph{SCAM}, 2014, pp.~215--224.

{[}6{]} T. Zimmermann, R. Premraj, and A. Zeller, ``Predicting defects
for Eclipse,'' in \emph{Proc. PROMISE}, 2007, pp.~9--20.

{[}7{]} E. Murphy-Hill, C. Parnin, and A. P. Black, ``How we refactor,
and how we know it,'' \emph{IEEE Trans. Softw. Eng.}, vol.~38, no. 1,
pp.~5--18, 2012.

{[}8{]} N. Nagappan and T. Ball, ``Use of relative code churn measures
to predict system defect density,'' in \emph{Proc. ICSE}, 2005,
pp.~284--292.

{[}9{]} A. E. Hassan, ``Predicting faults using the complexity of code
changes,'' in \emph{Proc. ICSE}, 2009, pp.~78--88.

{[}10{]} L. P. Hattori and M. Lanza, ``On the nature of commits,'' in
\emph{Proc. ASE Workshops}, 2008, pp.~63--71.

{[}11{]} M. Shahzad, M. Z. Shafiq, and A. X. Liu, ``A Large Scale
Exploratory Analysis of Software Vulnerability Life Cycles,'' in
\emph{Proc. ICSE}, 2012, pp.~771--781.

{[}12{]} G. Bavota, A. Qusef, R. Oliveto, A. De Lucia, and D. W.
Binkley, ``Are test smells really harmful? An empirical study,''
\emph{Empir. Softw. Eng.}, vol.~20, no. 4, pp.~1052--1094, 2015.

{[}13{]} D. G. Kleinbaum and M. Klein, \emph{Survival Analysis: A
Self-Learning Text}, 3rd ed.~Springer, 2012.

{[}14{]} B. Fluri, M. Wuersch, M. Pinzger, and H. C. Gall, ``Change
distilling: Tree differencing for fine-grained source code change
extraction,'' \emph{IEEE Trans. Softw. Eng.}, vol.~33, no. 11,
pp.~725--743, 2007.

{[}15{]} J. Y. Gil and G. Lalouche, ``When do software complexity
metrics mean nothing? --- When examined out of context,'' \emph{J.
Object Technology}, vol.~15, no. 1, 2016.

{[}16{]} J. C. Munson and S. G. Elbaum, ``Code churn: A measure for
estimating the impact of code change,'' in \emph{Proc. ICSM}, 1998,
pp.~24--31.

{[}17{]} F. Khomh, M. Di Penta, Y.-G. Guéhéneuc, and G. Antoniol, ``An
exploratory study of the impact of antipatterns on class change- and
fault-proneness,'' \emph{Empir. Softw. Eng.}, vol.~17, no. 3,
pp.~243--275, 2012.

{[}18{]} P. Kruchten, R. L. Nord, and I. Ozkaya, ``Technical debt: From
metaphor to theory and practice,'' \emph{IEEE Software}, vol.~29, no. 6,
pp.~18--21, 2012.

{[}19{]} A. Tornhill, \emph{Software Design X-Rays: Fix Technical Debt
with Behavioral Code Analysis}. Pragmatic Bookshelf, 2018.

{[}20{]} J. W. Ratcliff and D. E. Metzener, ``Pattern Matching: The
Gestalt Approach,'' \emph{Dr.~Dobb's Journal}, vol.~13, no. 7,
pp.~46--51, 1988. (Implemented in Python's standard library as
\texttt{difflib.SequenceMatcher.ratio()}.)

{[}21{]} L. R. Dice, ``Measures of the Amount of Ecologic Association
Between Species,'' \emph{Ecology}, vol.~26, no. 3, pp.~297--302, 1945.

{[}22{]} J. Śliwerski, T. Zimmermann, and A. Zeller, ``When do changes
induce fixes?'' in \emph{Proc. MSR Workshop at ICSE}, 2005, pp.~1--5.

{[}23{]} S. Kim, T. Zimmermann, K. Pan, and E. J. Whitehead Jr.,
``Automatic identification of bug-introducing changes,'' in \emph{Proc.
21st IEEE/ACM Int. Conf. on Automated Software Engineering (ASE)}, 2006,
pp.~81--90.

{[}24{]} H. W. Kuhn, ``The Hungarian Method for the Assignment
Problem,'' \emph{Naval Research Logistics Quarterly}, vol.~2, no. 1--2,
pp.~83--97, 1955.

{[}25{]} D. Spinellis, P. Louridas, and M. Kechagia, ``Software
evolution: the lifetime of fine-grained elements,'' \emph{PeerJ Computer
Science}, vol.~7, p.~e372, 2021.

{[}26{]} M. M. Lehman, ``Programs, life cycles, and laws of software
evolution,'' \emph{Proceedings of the IEEE}, vol.~68, no. 9,
pp.~1060--1076, 1980.

{[}27{]} D. L. Parnas, ``Software aging,'' in \emph{Proc. 16th
International Conference on Software Engineering (ICSE)}, 1994,
pp.~279--287.

{[}28{]} S. G. Eick, T. L. Graves, A. F. Karr, J. S. Marron, and A.
Mockus, ``Does code decay? Assessing the evidence from change management
data,'' \emph{IEEE Transactions on Software Engineering}, vol.~27, no.
1, pp.~1--12, 2001.

{[}29{]} P. Sentas, L. Angelis, and I. Stamelos, ``A statistical
framework for analyzing the duration of software projects,''
\emph{Empirical Software Engineering}, vol.~13, no. 2, pp.~147--184,
2008.

{[}30{]} G. Scanniello, ``Source code survival with the Kaplan Meier,''
in \emph{Proc. 27th IEEE International Conference on Software
Maintenance (ICSM)}, 2011, pp.~524--527.

{[}31{]} M. Claes, T. Mens, R. Di Cosmo, and J. Vouillon, ``A historical
analysis of Debian package incompatibilities,'' in \emph{Proc. 12th
Working Conference on Mining Software Repositories (MSR)}, 2015,
pp.~212--223.

{[}32{]} M. Goeminne and T. Mens, ``Towards a survival analysis of
database framework usage in Java projects,'' in \emph{Proc. 31st IEEE
International Conference on Software Maintenance and Evolution (ICSME)},
2015, pp.~551--555.

{[}33{]} X. Zheng, D. Zeng, H. Li, and F. Wang, ``Analyzing open-source
software systems as complex networks,'' \emph{Physica A: Statistical
Mechanics and its Applications}, vol.~387, no. 24, pp.~6190--6200, 2008.

{[}34{]} T. Sharma, M. Fragkoulis, and D. Spinellis, ``Does your
configuration code smell?'' in \emph{Proc. 13th Working Conference on
Mining Software Repositories (MSR)}, 2016, pp.~189--200.

{[}35{]} M. V. Mäntylä and C. Lassenius, ``Subjective evaluation of
software evolvability using code smells: an empirical study,''
\emph{Empirical Software Engineering}, vol.~11, no. 3, pp.~395--431,
2006.

\begin{center}\rule{0.5\linewidth}{0.5pt}\end{center}

\hypertarget{appendix-a.-full-repository-catalog}{%
\subsection{Appendix A. Full Repository
Catalog}\label{appendix-a.-full-repository-catalog}}

\textbf{Table A1.} All 120 GitHub repositories included in the CLSA
dataset, sorted by total lines mined in descending order. \emph{Age} =
months between the first and last observed commit; \emph{Lines Mined} =
TypeScript lines added during the observation period; \emph{Deaths} =
permanently deleted lines (hard delete + file delete); \emph{Survived} =
lines still alive at the snapshot date. \emph{Deaths} + \emph{Survived}
\textless{} \emph{Lines Mined} for most repositories; the remainder are
lines transformed via migration or modification
(\texttt{line\_evolution}) and are right-censored in the survival model.

\begin{longtable}[]{@{}
  >{\raggedleft\arraybackslash}p{(\columnwidth - 12\tabcolsep) * \real{0.0625}}
  >{\raggedright\arraybackslash}p{(\columnwidth - 12\tabcolsep) * \real{0.5208}}
  >{\raggedleft\arraybackslash}p{(\columnwidth - 12\tabcolsep) * \real{0.0833}}
  >{\raggedleft\arraybackslash}p{(\columnwidth - 12\tabcolsep) * \real{0.0833}}
  >{\raggedleft\arraybackslash}p{(\columnwidth - 12\tabcolsep) * \real{0.0833}}
  >{\raggedleft\arraybackslash}p{(\columnwidth - 12\tabcolsep) * \real{0.0833}}
  >{\raggedleft\arraybackslash}p{(\columnwidth - 12\tabcolsep) * \real{0.0833}}@{}}
\toprule\noalign{}
\begin{minipage}[b]{\linewidth}\raggedleft
\#
\end{minipage} & \begin{minipage}[b]{\linewidth}\raggedright
Repository
\end{minipage} & \begin{minipage}[b]{\linewidth}\raggedleft
Age (Mo.)
\end{minipage} & \begin{minipage}[b]{\linewidth}\raggedleft
Commits
\end{minipage} & \begin{minipage}[b]{\linewidth}\raggedleft
Lines Mined
\end{minipage} & \begin{minipage}[b]{\linewidth}\raggedleft
Deaths
\end{minipage} & \begin{minipage}[b]{\linewidth}\raggedleft
Survived
\end{minipage} \\
\midrule\noalign{}
\endhead
\bottomrule\noalign{}
\endlastfoot
1 & microsoft/\discretionary{}{}{}vscode & 126 & 118,215 & 4,031,373 &
966,838 & 1,909,391 \\
2 & Expensify/App & 35 & 99,575 & 2,835,458 & 1,039,257 & 1,124,026 \\
3 & twentyhq/\discretionary{}{}{}twenty & 41 & 10,201 & 2,170,359 &
785,557 & 990,896 \\
4 & DefinitelyTyped/\discretionary{}{}{}DefinitelyTyped & 163 & 69,085 &
1,853,038 & 724,178 & 773,375 \\
5 & ag-grid/\discretionary{}{}{}ag-grid & 130 & 23,141 & 1,570,216 &
539,199 & 693,573 \\
6 & microsoft/\discretionary{}{}{}FluidFramework & 116 & 16,807 &
1,468,504 & 558,400 & 569,540 \\
7 & remotion-dev/\discretionary{}{}{}remotion & 71 & 18,733 & 1,349,990
& 328,921 & 614,026 \\
8 & tamagui/\discretionary{}{}{}tamagui & 67 & 11,132 & 1,132,232 &
688,627 & 285,154 \\
9 & supabase/\discretionary{}{}{}supabase & 67 & 15,808 & 1,072,324 &
355,801 & 541,225 \\
10 & triggerdotdev/\discretionary{}{}{}trigger.\discretionary{}{}{}dev &
41 & 4,979 & 1,019,397 & 629,277 & 329,335 \\
11 & calcom/cal.com & 62 & 11,836 & 1,018,977 & 489,734 & 385,490 \\
12 & novuhq/novu & 57 & 10,841 & 1,013,499 & 346,998 & 479,665 \\
13 & medusajs/\discretionary{}{}{}medusa & 64 & 5,638 & 996,357 &
332,859 & 557,847 \\
14 & RocketChat/\discretionary{}{}{}Rocket.\discretionary{}{}{}Chat & 74
& 7,003 & 716,970 & 152,812 & 448,592 \\
15 & signalapp/\discretionary{}{}{}Signal-Desktop & 97 & 6,464 & 644,055
& 132,745 & 356,162 \\
16 & baidu/amis & 80 & 8,605 & 608,391 & 98,661 & 431,140 \\
17 & apache/\discretionary{}{}{}superset & 91 & 5,501 & 511,131 &
100,488 & 345,305 \\
18 & formatjs/\discretionary{}{}{}formatjs & 83 & 1,755 & 481,811 &
359,459 & 49,517 \\
19 & invoke-ai/\discretionary{}{}{}InvokeAI & 44 & 7,573 & 445,050 &
198,009 & 135,826 \\
20 & langchain-ai/\discretionary{}{}{}langchainjs & 39 & 4,113 & 426,748
& 164,586 & 226,707 \\
21 & Chocobozzz/\discretionary{}{}{}PeerTube & 122 & 6,943 & 385,050 &
77,413 & 193,259 \\
22 & cypress-io/\discretionary{}{}{}cypress & 108 & 4,078 & 380,500 &
141,923 & 186,742 \\
23 & chakra-ui/\discretionary{}{}{}chakra-ui & 81 & 5,407 & 352,584 &
149,979 & 135,123 \\
24 & scalar/scalar & 34 & 2,707 & 347,071 & 99,504 & 210,573 \\
25 & pnpm/pnpm & 116 & 5,701 & 282,313 & 51,632 & 155,069 \\
26 & angular/\discretionary{}{}{}angular-cli & 132 & 8,932 & 270,373 &
131,817 & 90,461 \\
27 & chakra-ui/ark & 43 & 2,117 & 261,237 & 77,226 & 130,846 \\
28 & dbeaver/\discretionary{}{}{}cloudbeaver & 73 & 3,765 & 241,048 &
69,158 & 101,436 \\
29 & gitpod-io/\discretionary{}{}{}gitpod & 67 & 4,763 & 239,729 &
88,651 & 107,528 \\
30 & outline/\discretionary{}{}{}outline & 54 & 4,593 & 234,788 & 37,591
& 163,326 \\
31 & remix-run/\discretionary{}{}{}react-router & 72 & 2,744 & 226,901 &
71,253 & 123,253 \\
32 & withastro/\discretionary{}{}{}astro & 62 & 7,425 & 213,448 & 93,000
& 79,477 \\
33 & streamlabs/\discretionary{}{}{}desktop & 102 & 3,796 & 203,891 &
59,700 & 103,844 \\
34 & bitpay/bitcore & 107 & 3,944 & 190,961 & 41,883 & 120,175 \\
35 & vuetifyjs/\discretionary{}{}{}vuetify & 105 & 4,944 & 147,520 &
49,036 & 69,644 \\
36 & nestjs/nest & 112 & 3,371 & 127,051 & 24,219 & 72,235 \\
37 & openkraken/\discretionary{}{}{}kraken & 31 & 1,489 & 120,970 &
15,239 & 93,066 \\
38 & colinhacks/zod & 74 & 1,459 & 120,755 & 53,009 & 51,065 \\
39 & freeCodeCamp/\discretionary{}{}{}freeCodeCamp & 60 & 2,672 &
115,248 & 29,261 & 69,556 \\
40 & ConsenSys-archive/\discretionary{}{}{}truffle & 61 & 3,016 &
109,911 & 23,941 & 62,897 \\
41 & Kanaries/Rath & 78 & 1,062 & 102,454 & 47,825 & 38,874 \\
42 & docmost/\discretionary{}{}{}docmost & 33 & 725 & 98,072 & 11,992 &
74,783 \\
43 & VSCodeVim/Vim & 125 & 3,573 & 97,979 & 22,830 & 43,825 \\
44 & chatboxai/\discretionary{}{}{}chatbox & 37 & 712 & 97,363 & 18,879
& 66,965 \\
45 & honojs/hono & 52 & 1,739 & 96,910 & 30,393 & 50,454 \\
46 & xuejianxianzun/\discretionary{}{}{}PixivBatchDownloader & 80 &
1,463 & 94,863 & 19,925 & 52,063 \\
47 & react-hook-form/\discretionary{}{}{}react-hook-form & 86 & 2,533 &
91,821 & 29,984 & 43,306 \\
48 & rjsf-team/\discretionary{}{}{}react-jsonschema-form & 80 & 534 &
89,148 & 19,354 & 54,360 \\
49 & radix-ui/\discretionary{}{}{}primitives & 65 & 950 & 87,511 &
33,537 & 34,849 \\
50 & verdaccio/\discretionary{}{}{}verdaccio & 82 & 682 & 84,750 &
31,712 & 36,699 \\
51 & kysely-org/\discretionary{}{}{}kysely & 62 & 1,003 & 82,445 &
14,102 & 50,790 \\
52 & jitsi/\discretionary{}{}{}jitsi-meet & 49 & 2,009 & 82,154 & 7,246
& 63,134 \\
53 & apollographql/\discretionary{}{}{}apollo-server & 119 & 2,001 &
81,388 & 47,825 & 19,591 \\
54 & video-dev/\discretionary{}{}{}hls.\discretionary{}{}{}js & 94 &
1,374 & 80,341 & 11,688 & 52,546 \\
55 & discordjs/\discretionary{}{}{}discord.\discretionary{}{}{}js & 113
& 2,672 & 73,397 & 14,132 & 45,367 \\
56 & alibaba/\discretionary{}{}{}lowcode-engine & 24 & 579 & 70,172 &
4,470 & 58,376 \\
57 & material-components/\discretionary{}{}{}material-components-web &
99 & 716 & 65,544 & 10,730 & 45,161 \\
58 & konvajs/konva & 135 & 732 & 65,479 & 21,623 & 38,112 \\
59 & xyflow/xyflow & 79 & 2,327 & 62,053 & 20,406 & 25,021 \\
60 & nanbingxyz/\discretionary{}{}{}5ire & 17 & 1,207 & 49,551 & 10,931
& 30,118 \\
61 & kanbn/kan & 30 & 560 & 48,634 & 6,846 & 35,186 \\
62 & clauderic/\discretionary{}{}{}dnd-kit & 66 & 761 & 47,312 & 17,571
& 20,664 \\
63 & yangshun/\discretionary{}{}{}tech-interview-handbook & 44 & 587 &
40,731 & 9,471 & 23,756 \\
64 & Koenkk/\discretionary{}{}{}zigbee2mqtt & 58 & 698 & 40,325 & 5,545
& 22,415 \\
65 & rrweb-io/rrweb & 89 & 796 & 38,904 & 4,697 & 29,249 \\
66 & shuding/nextra & 53 & 710 & 37,259 & 13,976 & 14,760 \\
67 & dexie/Dexie.js & 132 & 983 & 36,085 & 6,710 & 21,546 \\
68 & adonisjs/core & 88 & 764 & 35,834 & 19,563 & 10,990 \\
69 & whyour/\discretionary{}{}{}qinglong & 62 & 1,254 & 33,088 & 7,596 &
18,817 \\
70 & Yoctol/\discretionary{}{}{}bottender & 27 & 255 & 26,580 & 2,099 &
21,411 \\
71 & gitbrent/\discretionary{}{}{}PptxGenJS & 83 & 1,107 & 23,696 &
8,566 & 6,394 \\
72 & gajus/slonik & 67 & 476 & 23,099 & 5,526 & 10,618 \\
73 & vadimdemedes/\discretionary{}{}{}ink & 86 & 267 & 23,070 & 1,766 &
18,794 \\
74 & preactjs/\discretionary{}{}{}signals & 45 & 511 & 23,035 & 3,900 &
16,296 \\
75 & jd-opensource/\discretionary{}{}{}taro-ui & 93 & 358 & 23,024 & 882
& 20,513 \\
76 & soybeanjs/\discretionary{}{}{}soybean-admin & 58 & 714 & 22,603 &
10,736 & 6,537 \\
77 & vercel/swr & 76 & 565 & 22,398 & 5,335 & 13,151 \\
78 & typestack/\discretionary{}{}{}routing-controllers & 114 & 360 &
21,451 & 5,662 & 7,029 \\
79 & sismo-core/\discretionary{}{}{}sismo-badges & 7 & 213 & 19,457 &
4,720 & 12,874 \\
80 & lokalise/\discretionary{}{}{}i18n-ally & 66 & 830 & 19,206 & 3,230
& 10,553 \\
81 & streamich/\discretionary{}{}{}react-use & 74 & 708 & 18,229 & 2,448
& 11,852 \\
82 & conventional-changelog/\discretionary{}{}{}commitlint & 84 & 220 &
17,461 & 1,072 & 10,290 \\
83 & nestjsx/crud & 55 & 225 & 15,154 & 5,359 & 7,556 \\
84 & seek-oss/\discretionary{}{}{}playroom & 87 & 134 & 14,078 & 2,699 &
9,210 \\
85 & steven-tey/\discretionary{}{}{}novel & 22 & 347 & 13,549 & 6,442 &
4,342 \\
86 & algolia/\discretionary{}{}{}docsearch & 73 & 305 & 12,648 & 1,430 &
9,708 \\
87 & pmndrs/zustand & 85 & 326 & 12,015 & 2,789 & 5,930 \\
88 & Lissy93/\discretionary{}{}{}web-check & 47 & 220 & 11,909 & 2,602 &
6,511 \\
89 & slab/quill & 37 & 212 & 11,820 & 691 & 9,731 \\
90 & Authenticator-Extension/\discretionary{}{}{}Authenticator & 80 &
340 & 10,883 & 2,227 & 5,753 \\
91 & appwrite/\discretionary{}{}{}appwrite & 71 & 46 & 8,904 & 5,073 &
3,808 \\
92 & Molunerfinn/\discretionary{}{}{}PicGo & 76 & 183 & 8,535 & 1,342 &
5,864 \\
93 & fkhadra/\discretionary{}{}{}react-toastify & 102 & 438 & 8,056 &
3,418 & 2,839 \\
94 & iptv-org/iptv & 32 & 165 & 7,533 & 2,458 & 2,848 \\
95 & shadcn-ui/\discretionary{}{}{}taxonomy & 6 & 82 & 7,221 & 765 &
5,258 \\
96 & astriaai/\discretionary{}{}{}headshots-starter & 19 & 196 & 7,197 &
1,234 & 4,949 \\
97 & nomcopter/\discretionary{}{}{}react-mosaic & 109 & 93 & 7,006 &
1,605 & 4,100 \\
98 & vercel/\discretionary{}{}{}next-learn & 67 & 126 & 6,555 & 1,573 &
3,566 \\
99 & postcss/\discretionary{}{}{}postcss & 128 & 359 & 6,346 & 596 &
4,188 \\
100 & saltyshiomix/\discretionary{}{}{}nextron & 94 & 313 & 5,965 &
2,523 & 1,831 \\
101 & Tencent/wujie & 47 & 163 & 5,533 & 423 & 4,561 \\
102 & haishanh/yacd & 42 & 132 & 5,522 & 978 & 3,750 \\
103 & improbable-eng/\discretionary{}{}{}grpc-web & 66 & 91 & 4,934 &
787 & 3,019 \\
104 & chanind/\discretionary{}{}{}hanzi-writer & 59 & 16 & 4,864 & 31 &
4,751 \\
105 & fastmail/\discretionary{}{}{}Squire & 40 & 81 & 4,786 & 101 &
4,519 \\
106 & unform/unform & 23 & 78 & 3,788 & 1,362 & 1,376 \\
107 & airbnb/\discretionary{}{}{}react-sketchapp & 28 & 58 & 3,485 & 387
& 2,657 \\
108 & unjs/\discretionary{}{}{}magic-regexp & 44 & 86 & 2,940 & 349 &
1,899 \\
109 & piotrwitek/\discretionary{}{}{}react-redux-typescript-guide & 60 &
96 & 2,788 & 644 & 1,301 \\
110 & markedjs/\discretionary{}{}{}marked & 34 & 101 & 2,623 & 480 &
1,568 \\
111 & formkit/\discretionary{}{}{}auto-animate & 46 & 106 & 2,550 & 208
& 2,020 \\
112 & cruip/\discretionary{}{}{}tailwind-landing-page-template & 22 & 10
& 2,155 & 464 & 1,232 \\
113 & alexjoverm/\discretionary{}{}{}typescript-library-starter & 24 &
80 & 2,024 & 1,274 & 201 \\
114 & unplugin/\discretionary{}{}{}unplugin-icons & 61 & 179 & 1,923 &
522 & 963 \\
115 & danilowoz/\discretionary{}{}{}react-content-loader & 84 & 32 &
1,811 & 358 & 1,162 \\
116 & Nutlope/\discretionary{}{}{}restorePhotos & 17 & 100 & 1,739 & 356
& 802 \\
117 & niklashigi/\discretionary{}{}{}apk-mitm & 54 & 93 & 1,419 & 249 &
863 \\
118 & babaohuang/\discretionary{}{}{}GeminiProChat & 20 & 78 & 1,256 &
163 & 637 \\
119 & catppuccin/\discretionary{}{}{}catppuccin & 40 & 29 & 627 & 167 &
364 \\
120 & SamKirkland/\discretionary{}{}{}FTP-Deploy-Action & 26 & 22 & 320
& 108 & 135 \\
\end{longtable}

\textbf{Total:} 120 repositories; 32,464,566 lines mined; 11,009,579
deaths; 14,990,228 survived.

\end{document}